\documentclass[twocolumn,preprintnumbers,superscriptaddress,landscape,nofootinbib]{revtex4}
\usepackage[utf8]{inputenc}
\usepackage{amsmath}
\usepackage{amsthm}
\usepackage{amssymb}
\usepackage{float}
\usepackage{braket}
\usepackage{graphicx}
\usepackage{url}
\usepackage[colorlinks=true, pdfstartview=FitV, linkcolor=red, citecolor=blue, urlcolor=blue]{hyperref}
\usepackage{slashed}
\usepackage[normalem]{ulem}
\graphicspath{{./figures/}}

\newcommand{\be}{\begin{equation}}      
\newcommand{\ee}{\end{equation}}      
\newcommand{\bea}{\begin{eqnarray}}      
\newcommand{\eea}{\end{eqnarray}}

\begin{document}

\title{Vortical Quantum Memory}
\author{Kazuki Ikeda}
\email[]{kazuki7131ATgmail.com}
\affiliation{Department of Mathematics and Statistics
$\&$ Centre for Quantum Topology and Its Applications (quanTA), University of Saskatchewan, Saskatoon, Saskatchewan S7N 5E6, Canada}

\author{Dmitri E. Kharzeev}
\email[]{dmitri.kharzeevATstonybrook.edu}

\affiliation{Center for Nuclear Theory, Department of Physics and Astronomy, Stony Brook University, Stony Brook, New York 11794-3800, USA}
\affiliation{Department of Physics, Brookhaven National Laboratory, Upton, New York 11973-5000}

\author{Yuta Kikuchi}
\email[]{ykikuchiATbnl.gov}
\affiliation{Department of Physics, Brookhaven National Laboratory, Upton, New York 11973-5000}

\bibliographystyle{unsrt}

\begin{abstract}

Quantum memory is a crucial component of a quantum information processor, just like a classical memory is a necessary ingredient of a conventional computer. Moreover, quantum memory of light would serve as a quantum repeater  needed for quantum communication networks. Here we propose to realize the quantum memory coupled to magnetic modes of electromagnetic radiation (e.g. those found in cavities and optical fibers) using vortices in Quantum Hall systems. We describe the response to an external magnetic mode as a quench, and find that it sets an oscillation between vortices and antivortices, with the period controlled by the amplitude of the magnetic mode, and the phase locked to the phase of the magnetic mode. We support our proposal by real-time Hamiltonian field theory simulations. 

\end{abstract}

\maketitle

\section{Introduction}

The development of quantum computers and quantum communications requires quantum memory that stores a quantum state for later retrieval. Of particular interest is a quantum memory that can couple directly to electromagnetic radiation. The main challenge in developing a robust quantum memory is overcoming noise and decoherence. This naturally draws attention to systems in which decoherence can be reduced to a minimum. Because of this, here we propose to consider Quantum Hall systems (realized in topological materials with a gapped bulk~\cite{qi2008topological,2008PhRvB..78s5125S,slager2013space,2017NatCo...8...50P}) -- with their topologically protected edge modes~\cite{PhysRevLett.71.3697,hatsugai1993edge,kawabata2018anomalous,susstrunk2015observation}, they offer a promise of long coherence time and reduced dissipation. Quantum Hall effect \cite{doi:10.1143/JPSJ.39.279,PhysRevB.23.5632,0022-3719-14-23-022,PhysRevLett.45.494} continuously finds new applications and new connections emerge, including its relevance for the Langlands program pointed out recently  \cite{IKEDA2018136,doi:10.1063/1.4998635,2018arXiv181211879I,matsuki2019comments,matsuki2021fractal}. 
\vskip0.3cm

Quantum Hall systems can be controlled by an external magnetic flux, such as the one created by magnetic modes in cavities and optical fibers. In this paper we will utilize a two-band model of a Quantum Hall system that describes vortices and antivortices that occur in the Brillouin zone of a Quantum Hall material. The ground state, depending on the value of an external magnetic flux, contains antivortices, or no vortices or antivortices at all. However when a rapid change in the external magnetic flux occurs (we will call this a ``topological quench", as it changes the topology of the system), the quantum state becomes a superposition of vortices and antivortices, with a phase determined by the moment of time when the quench took place\footnote{Of course, the value of magnetic flux after the quench has to allow for the existence of vortices in the system.}. As we will show below, the period of quantum oscillations between vortices and antivortices is determined by the amplitude of the external magnetic flux, and varies in a very wide interval. Therefore, both the phase and the amplitude of an electromagnetic wave are recorded in the Quantum Hall state, which thus makes it a quantum memory. 

\vskip0.3cm
Traditionally, quantum Hall effect (QHE) requires an external magnetic field. The QHE however appears even without an external magnetic field in magnetic topological insulators. A topological insulator is a material that exhibits a metallic state on the surface while the interior is in an insulating state. A magnetic topological insulator can be created through a proximity effect due to the presence of magnetic dopants, or it can be a ferromagnetic material with internal magnetization. The interplay between the metallic state of the surface and ferromagnetism generates the quantum Hall effect even in the absence of a magnetic field~\cite{2013Sci...340..167C,2015arXiv150807106L}. This phenomenon is called the quantum anomalous Hall (QAH) effect, which is distinct from the integer quantum Hall effect, because edge currents can be generated without the application of an external magnetic field.

\vskip0.3cm
It is tempting to think of this topological quantum system with periodic oscillations between vortices and antivortices in terms of a ``topological time crystal", but our quantum state is {\it not} a ground state of the Quantum Hall system at a given value of an external magnetic flux. Instead, it is an excited state formed by the topological quench, and the time evolution of this quantum state is controlled by the parameters of the quench. 
\vskip0.3cm
The paper is organized as follows. In Section \ref{sec:quench} we describe the two-band model of the Quantum Anomalous Hall systems that we use, discuss the vortices and antivortices that it contains, and outline the topological quench protocol that we use in our real-time Hamiltonian field theory simulation. In Section \ref{sec:memory} we present the results of our real-time simulation, and show that the parameters of topological quench are indeed encoded in the phase and period of quantum vortex-antivortex oscillation that we observe. Finally, in Section \ref{sec:discussion} we discuss the results and present an outlook for future studies.

\section{Topological Quench in Quantum Hall System}\label{sec:quench}
\subsection{The two-band model}
Here we consider the quantum anomalous Hall effect, which occurs when the external magnetic field is zero but the spins are polarized. This model is described by the following simple two-band Hamiltonian~\cite{PhysRevB.74.085308}
\begin{align}
\begin{aligned}
        H&=\sum_{i}R_i(m,k_x,k_y)\sigma_i, \\
        R_x&=\sin k_x,R_y=\sin k_y, R_z=m-\cos k_x-\cos k_y .
\end{aligned}
\end{align}
The physical interpretation of this model in terms of the relevant degrees of freedom depends on it material realization -- for example, the two bands can represent either different spin states of electrons with a spin-orbit interaction, or orbital degrees of freedom with band hybridization. 

\begin{figure}[H]
\begin{minipage}{0.49\hsize}
                \centering
    \includegraphics[width=\hsize]{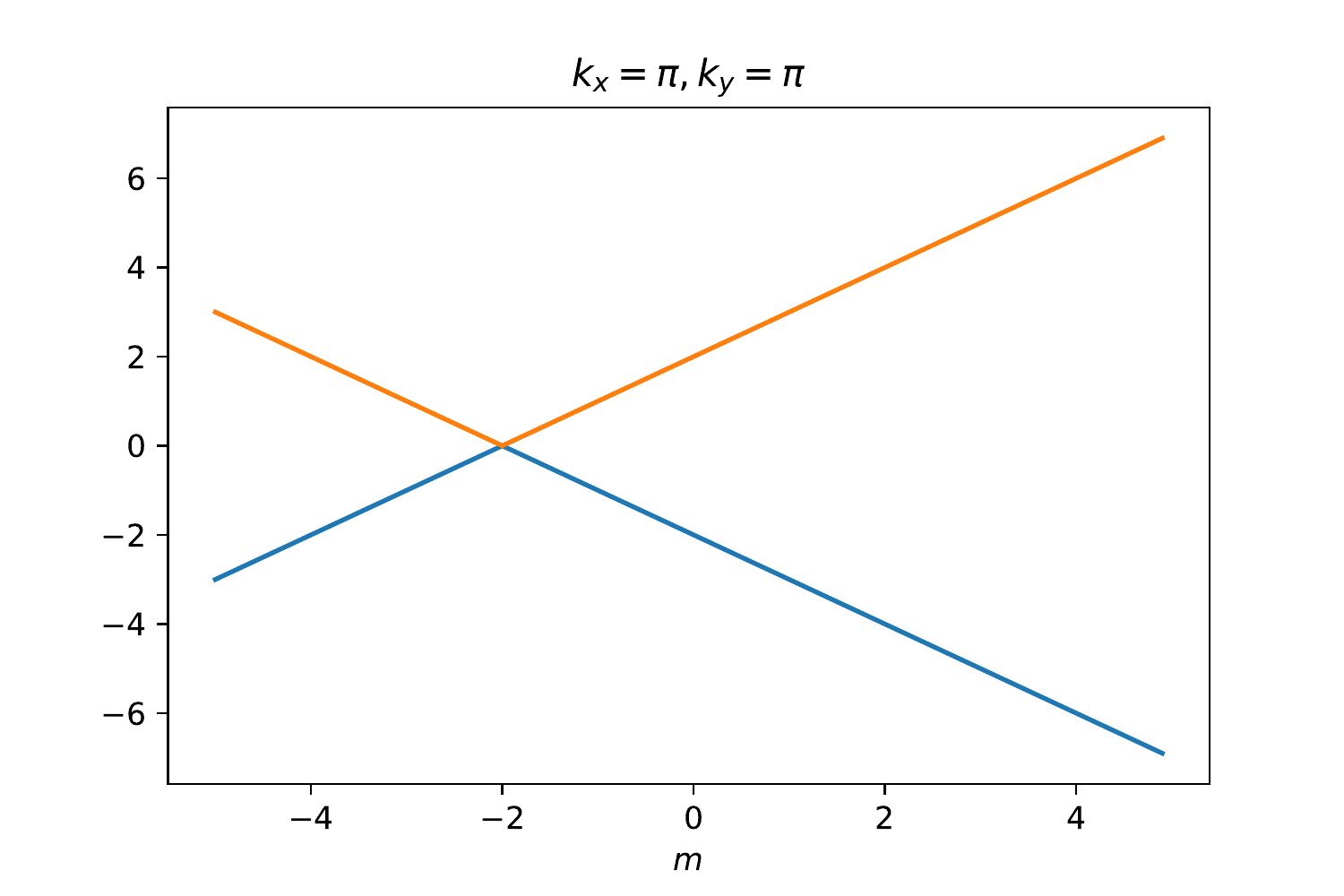}
\end{minipage}
\begin{minipage}{0.49\hsize}
                \centering
    \includegraphics[width=\hsize]{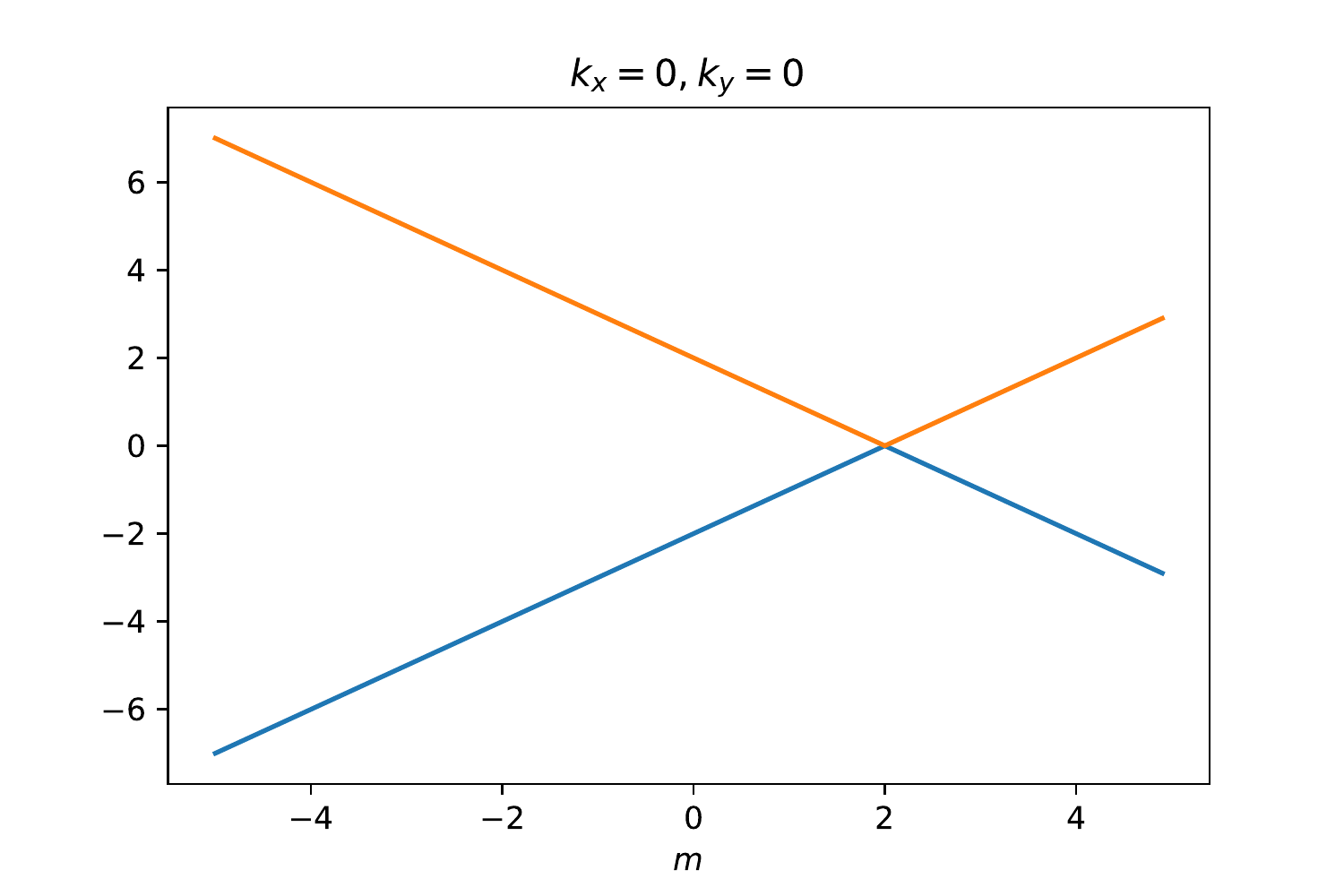}
\end{minipage}
    \caption{Energy eigenvalues at $k_x=k_y=\pi$ and $0$.}
    \label{fig:energy}
\end{figure}

This model has the high symmetry points $k_i=0,\pi$, i.e. momentum points at the edges and center that represent all combinations of $k_i=0,\pi$ for each direction of $k_i$: $(0,0),(\pi,0),(0,\pi),(\pi,\pi)$. As plotted in Fig.\ref{fig:energy}, there is a band crossing point at $m=-2$ if $k_x=k_y=\pi$, at which the energy gap decreases as $m$ approaches -2. These are the points around which the vortices of Berry curvature are localized. As we will explain in detail below, we construct the topological invariant of these vortices by assigning the $\mathbb{Z}_2$-index to the vortex based on its rotation direction : clockwise (-1) or anticlockwise (+1).

The lower band has the eigenvalue and the eigenstate 
\begin{align}
    E(k)&=-\sqrt{R_x(k)^2+R_y(k)^2+R_z(k)^2}\\
    \ket{\Psi^A(k)}&=\frac{1}{N_A}
    \begin{pmatrix}
    R_z(k)-|R(k)|\\
    R_x(k)-iR_y(k)
    \end{pmatrix},
\end{align}
where $N_A$ is the normalization factor. This wave function becomes ill-defined if $R_z \ge0$ and $R_x(k)=R_y(k)=0$. In our case,  this happens for example at $k_x=k_y=\pi$, $m>-2$. 

There is another way to express the same state by shifting the phase 
\begin{equation}
    \ket{\Psi^B(k)}=\frac{\frac{R_z+|R|}{R_x+iR_y}}{\bigg|\frac{R_z+|R|}{R_x+iR_y}\bigg|}\ket{\Psi^A(k)}=\frac{1}{N_B}
    \begin{pmatrix}
    -R_x(k)+iR_y(k)\\
    R_z(k)+|R(k)|
    \end{pmatrix},
\end{equation}
which is well-defined at $R_z >0$ and $R_x(k)=R_y(k)=0$. This wave function becomes ill-defined if $R_z \le0$ and $R_x(k)=R_y(k)=0$. In our case,  this happens for example at $k_x=k_y=0$ of $m<2$. 

\subsection{Vortices and anti-vortices}

Because the wave function is ill-defined at some points in the Brillouin zone, we need to remove these points by cutting the Brillouin zone into two and using two different representations of the wave function. These representations are connected by a gauge transformation 
\begin{equation}
    \ket{\Psi^B}=e^{i\phi(k)}\ket{\Psi^A}. 
\end{equation}
The Berry connection
\begin{equation}
 \mathcal{A}^l(k)=-i\bra{\psi^l(k)}d\psi^l(k)\rangle,   
\end{equation}
is transformed as 
\begin{equation}
    \mathcal{A}^B=\mathcal{A}^A+\nabla_k\phi(k). 
\end{equation}
Let us discuss the structure of the ground state for different values of magnetic flux.
\begin{itemize}
\item{
For $m>2$, $R_z(k)$ is always positive for all the four high symmetry points. So we can use $\Psi^B(k)$ for the entire Brillouin zone. Therefore the first Chern number
\begin{equation}
  C=\frac{1}{2\pi}\int_{T_\text{BZ}}d^2k d\mathcal{A}^B  
\end{equation}
is zero, 
$C=0$, and no vortices and antivortices exist.}

\item{
For $0<m<2$, $R_z(k)$ is negative at $(0,0)$ and positive at the other three points $(\pi,\pi), (0,\pi), (\pi,0)$. So we can choose $\Psi^A$ around a small disk of $(0,0)$ and $\Psi^B$ outside of it. By taking the disk sufficiently small, we find $\phi(k)=-\theta$ and $C=\frac{1}{2\pi}\int_{0}^{2\pi}d\theta \partial_\theta \phi=-1$, which corresponds to an antivortex. 
}

\item{
For $-2<m<0$, $R_z(k)$ is positive at $(\pi,\pi)$ and negative at the other three points $(\pi,\pi), (0,\pi), (\pi,0)$. So we can choose $\Psi^B$ around a small disk of $(\pi,\pi)$ and $\Psi^A$ outside of it. By repeating the same argument, we again find $C=-1$.
}

\item{
For $m<-2$, $R_z(k)$ is negative for all the four special points. So we can use $\Psi^A(k)$ for the entire Brillouin zone. Therefore $C=0$. 
}

\item{
We cannot define the Chern number for $m=0, \pm 2$, where the gap closes.
}
\end{itemize}

The Hall conductance associated with the $l$-th band is explained by the Kubo formula \cite{doi:10.1143/JPSJ.12.570,1982PhLA...90..474W,0022-3719-15-22-005,PhysRevLett.49.405}
\begin{align}
\begin{aligned}
        \sigma^l_{xy}&=-\frac{e^2}{h}\frac{1}{2\pi}\int_{T_{\text{BZ}}}d^2k d\mathcal{A}^l= -\frac{e^2}{h}\ C
\end{aligned}    
\end{align}
where $\psi^l(k)$ is the $l$-th Bloch function. We consider a quench process of $\psi^l$
\begin{equation}
    \ket{\psi^l(t,k)}=e^{-it H'}\ket{\psi^l(k)}, 
\end{equation}
where $H'=H(m')$, and then we find the time-dependent Berry connection,
\begin{equation}
    \mathcal{A}^l(t,k)=-i\bra{\psi^l(t,k)}d\psi^l(t,k)\rangle.
\end{equation}

\begin{figure*}
\begin{minipage}{0.32\hsize}
                \centering
    \includegraphics[width=\hsize]{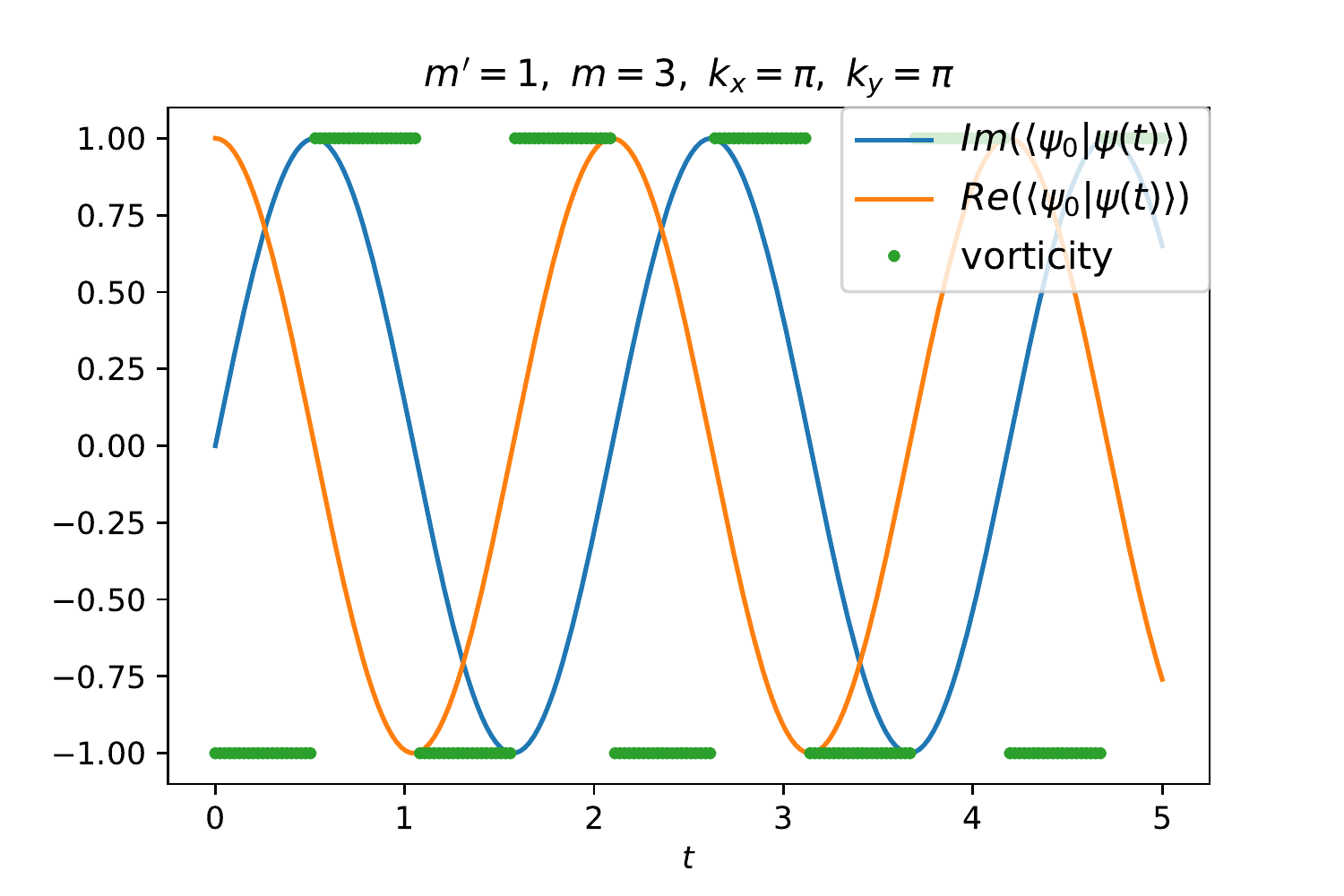}
\end{minipage}
\begin{minipage}{0.32\hsize}
                \centering
    \includegraphics[width=\hsize]{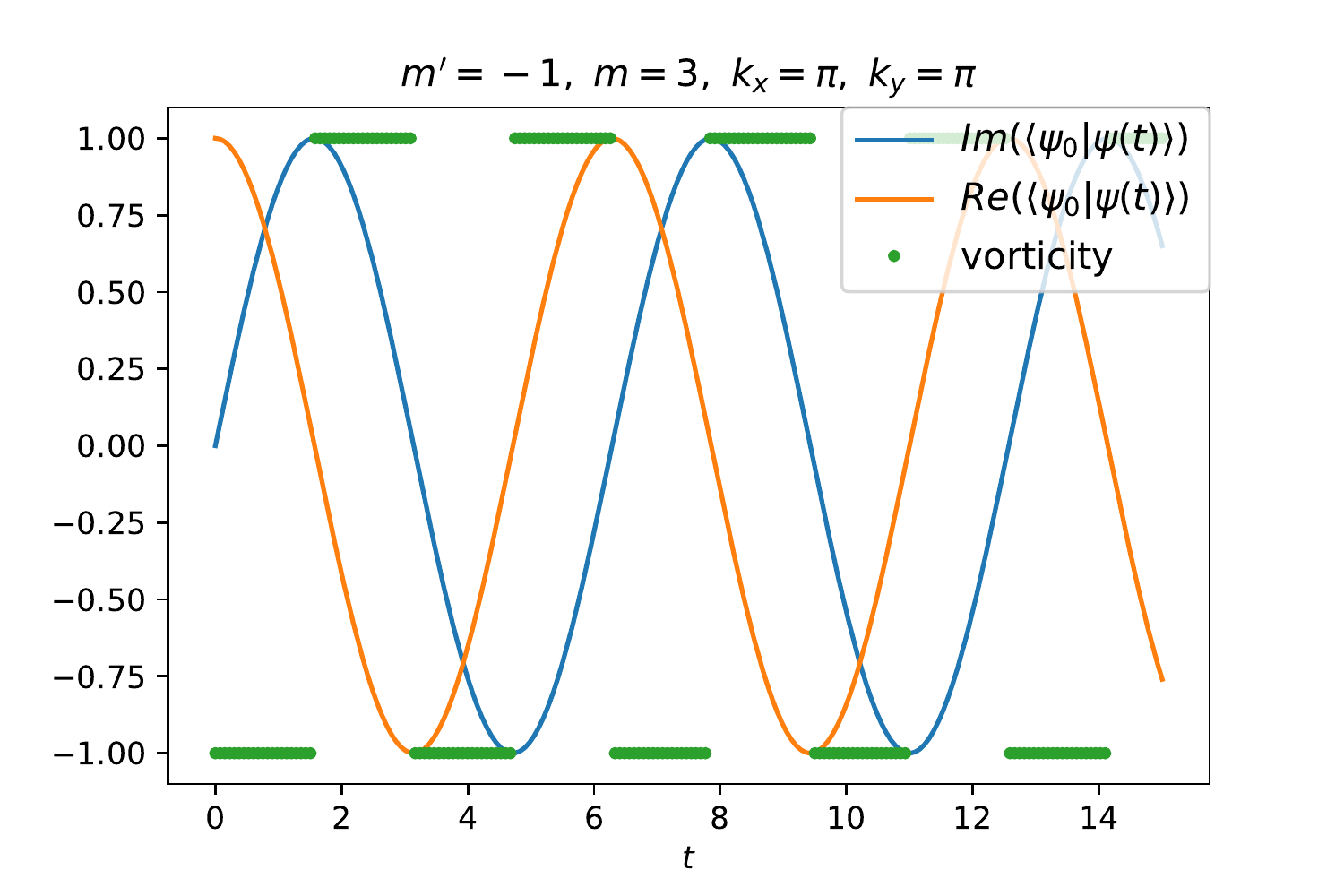}
\end{minipage}
\begin{minipage}{0.32\hsize}
                \centering
    \includegraphics[width=\hsize]{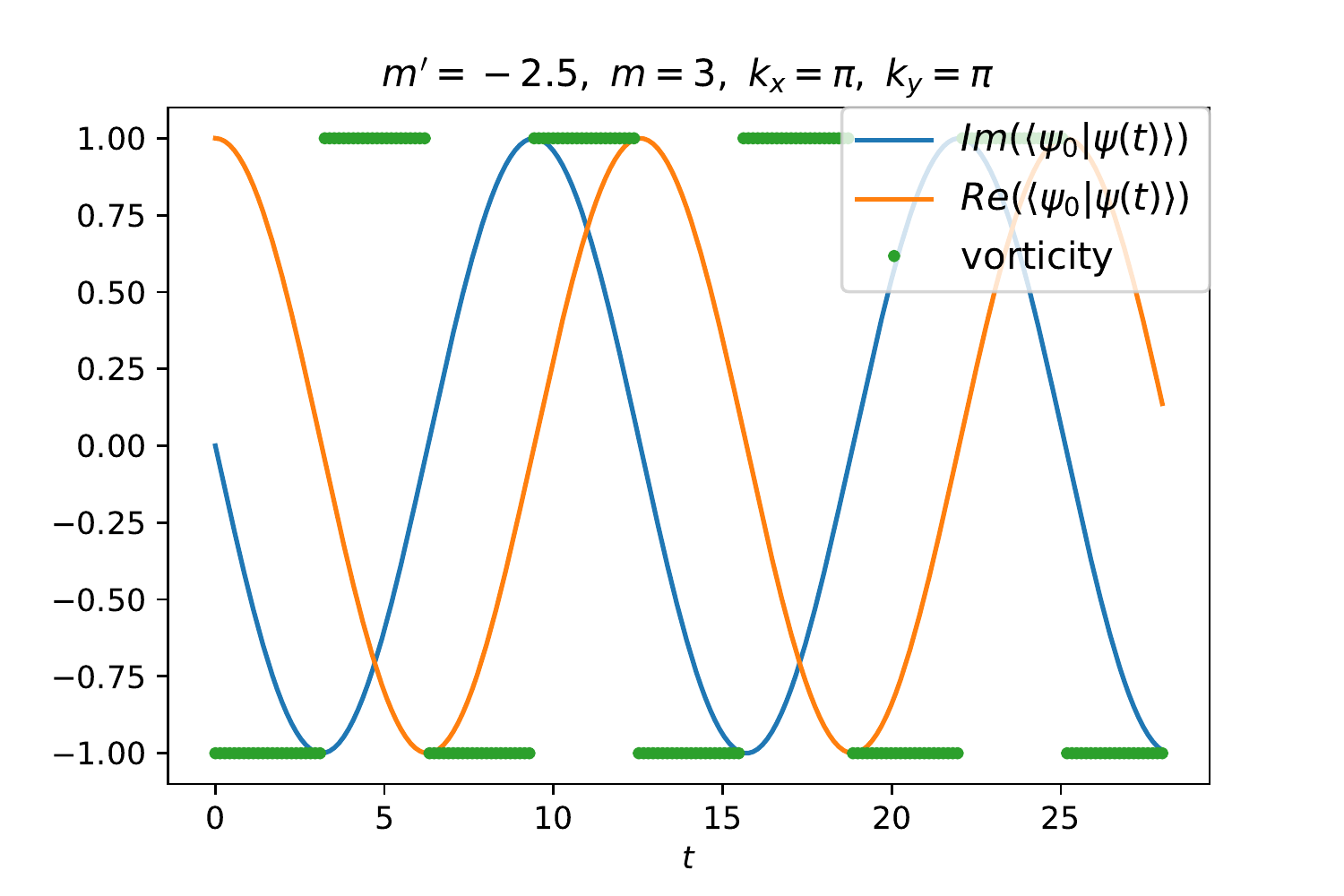}
\end{minipage}
    \caption{Mass dependence of vorticity change near $k_x=k_y=\pi$. [Left] $m'=1.0$, [Middle] $m'=-1.0$, [Right] $m'=-2.5$. }
    \label{fig:pi}
\end{figure*}

\begin{figure*}
\begin{minipage}{0.32\hsize}
                \centering
    \includegraphics[width=\hsize]{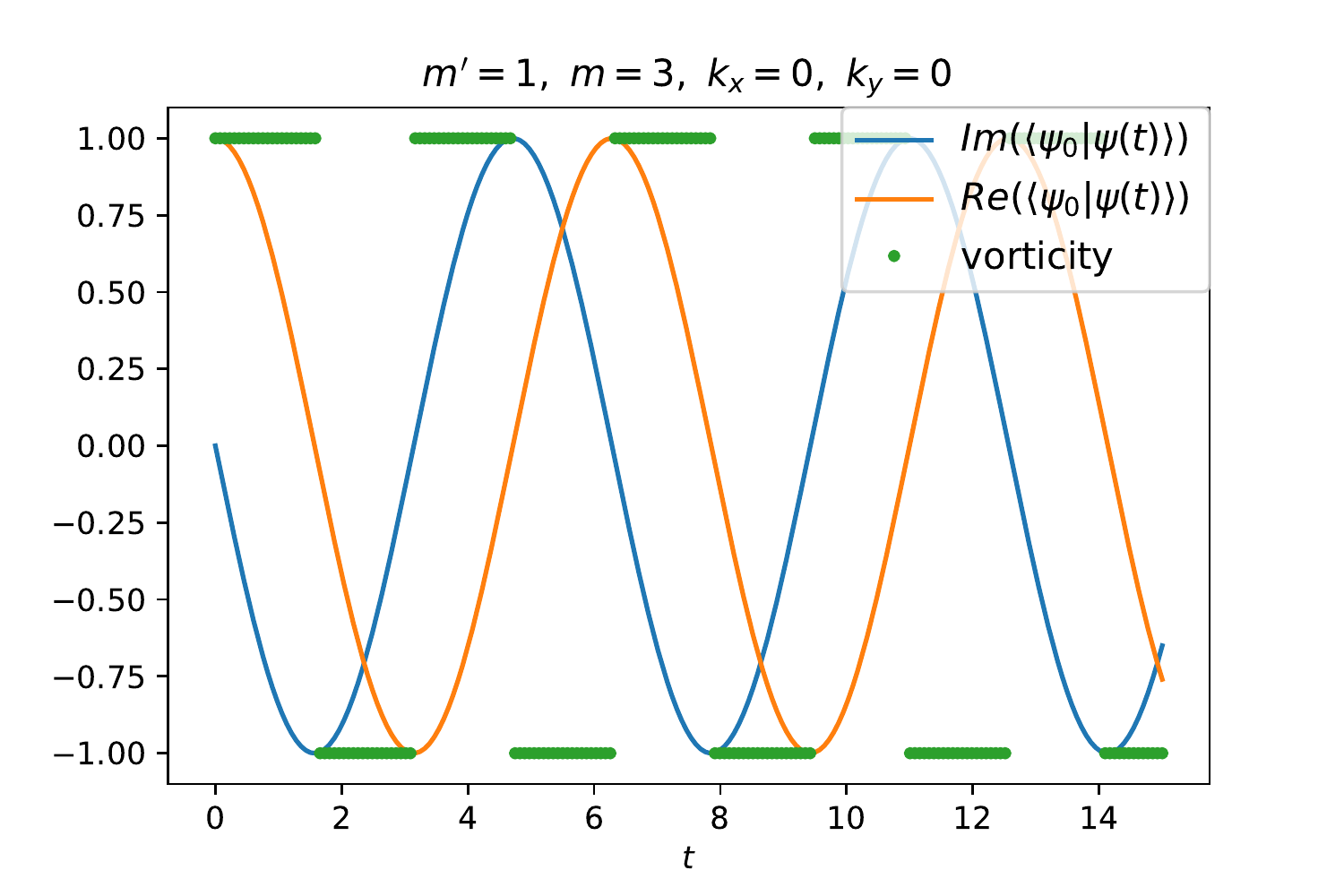}
\end{minipage}
\begin{minipage}{0.32\hsize}
                \centering
    \includegraphics[width=\hsize]{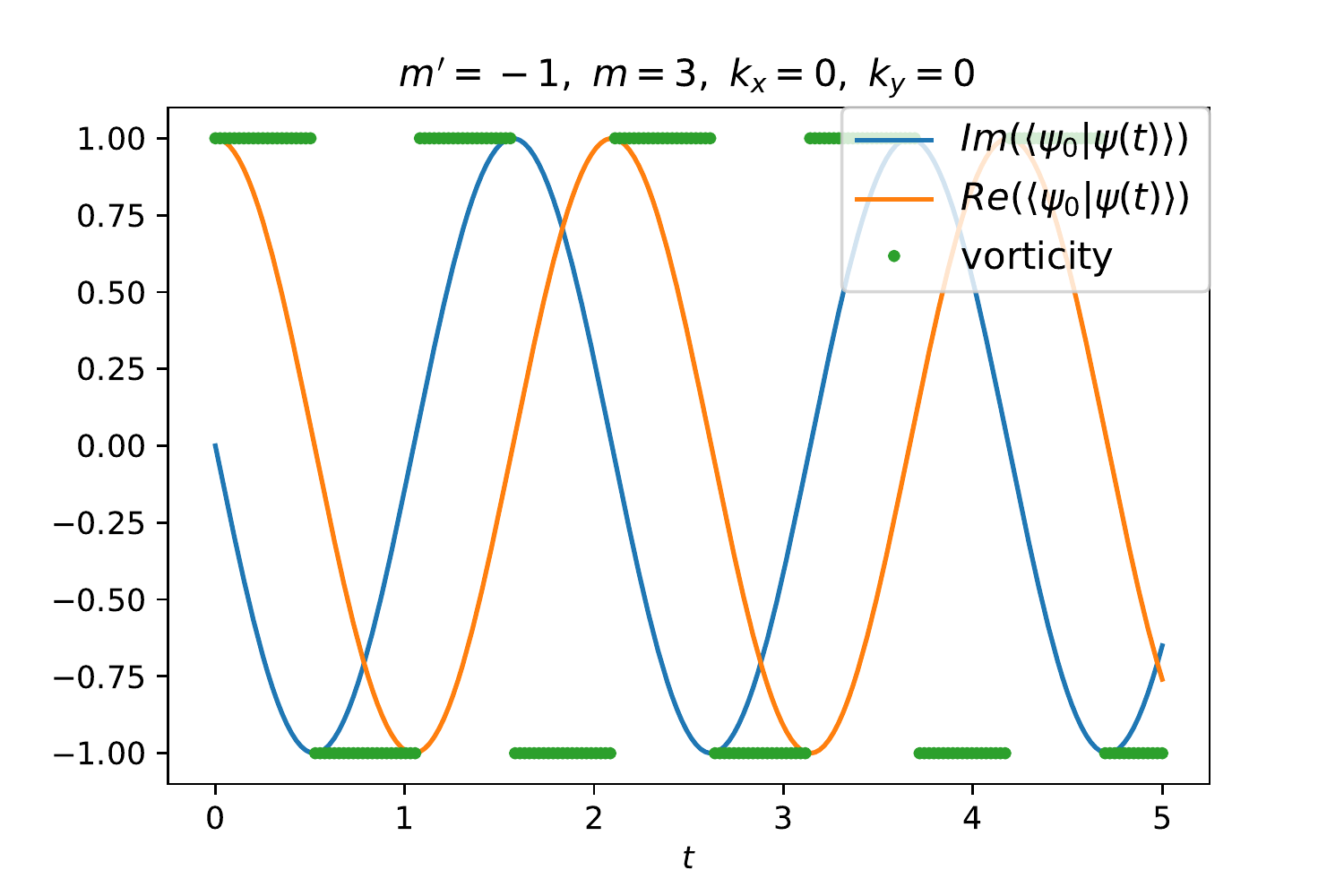}
\end{minipage}
\begin{minipage}{0.32\hsize}
                \centering
    \includegraphics[width=\hsize]{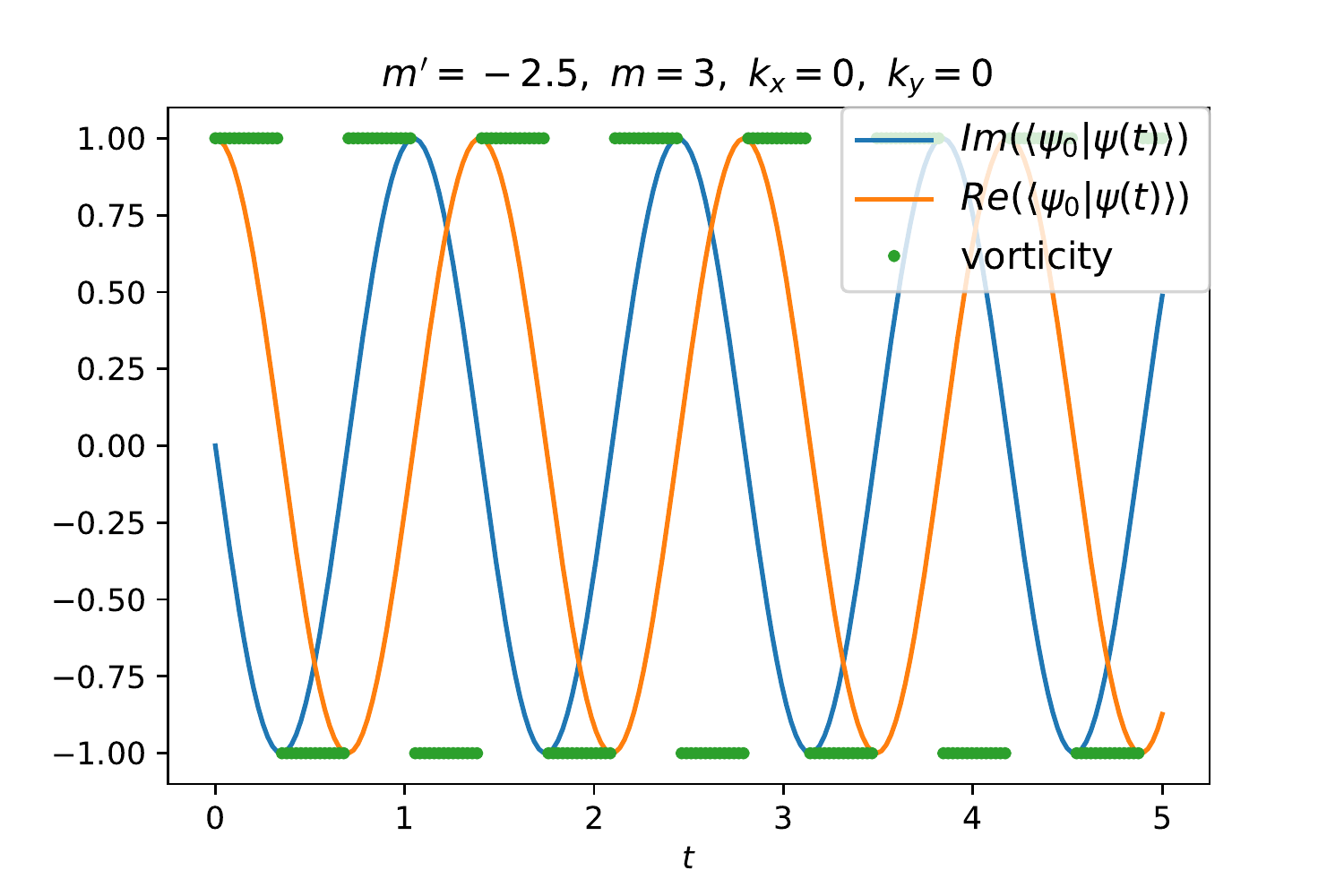}
\end{minipage}
    \caption{Mass dependence of vorticity change near $k_x=k_y=0$. [Left] $m'=1.0$, [Middle] $m'=-1.0$, [Right] $m'=-2.5$. }
    \label{fig:zero}
\end{figure*}

\begin{figure*}
\begin{minipage}{0.49\hsize}
            \centering
    \includegraphics[width=\hsize]{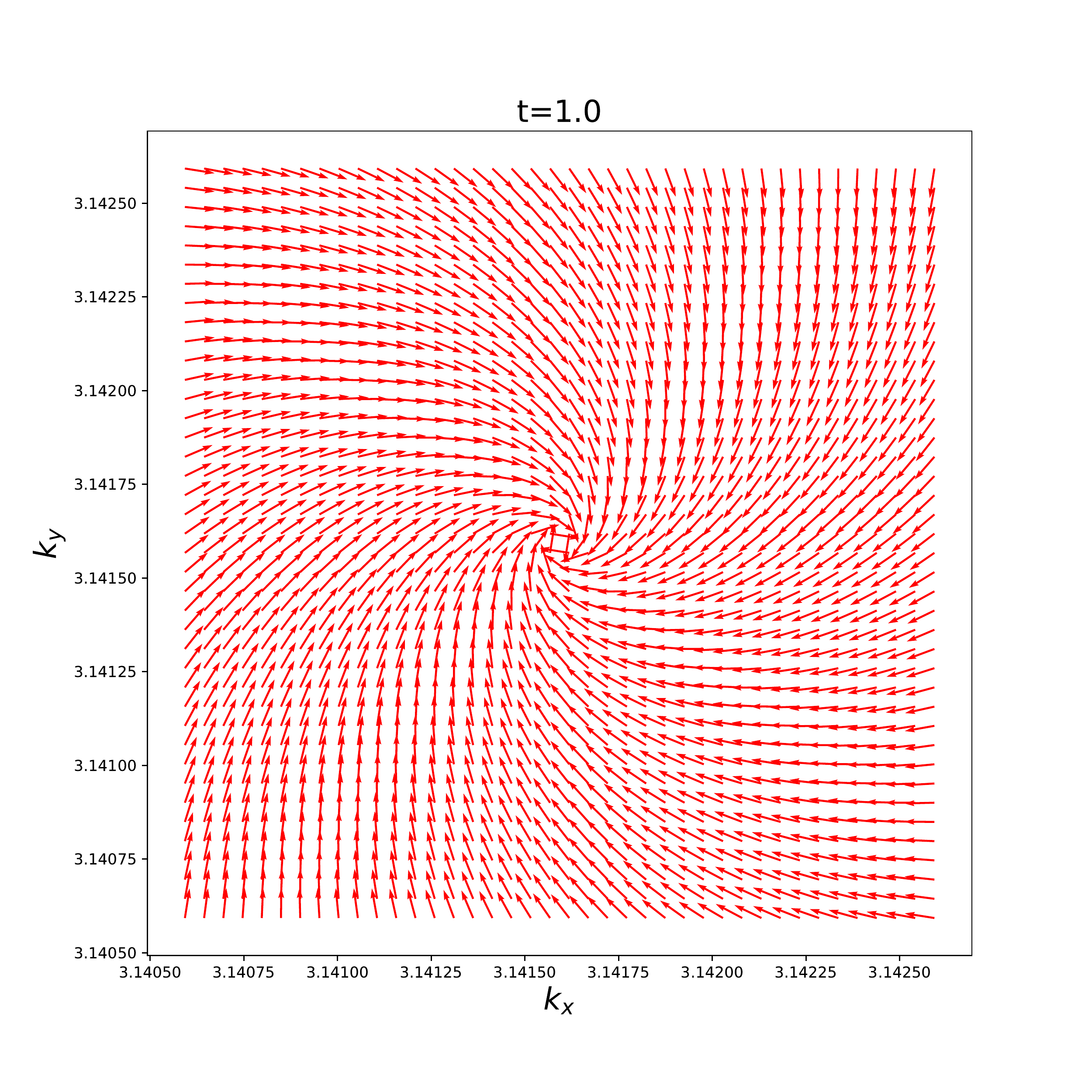}
\end{minipage}
\begin{minipage}{0.49\hsize}
            \centering
    \includegraphics[width=\hsize]{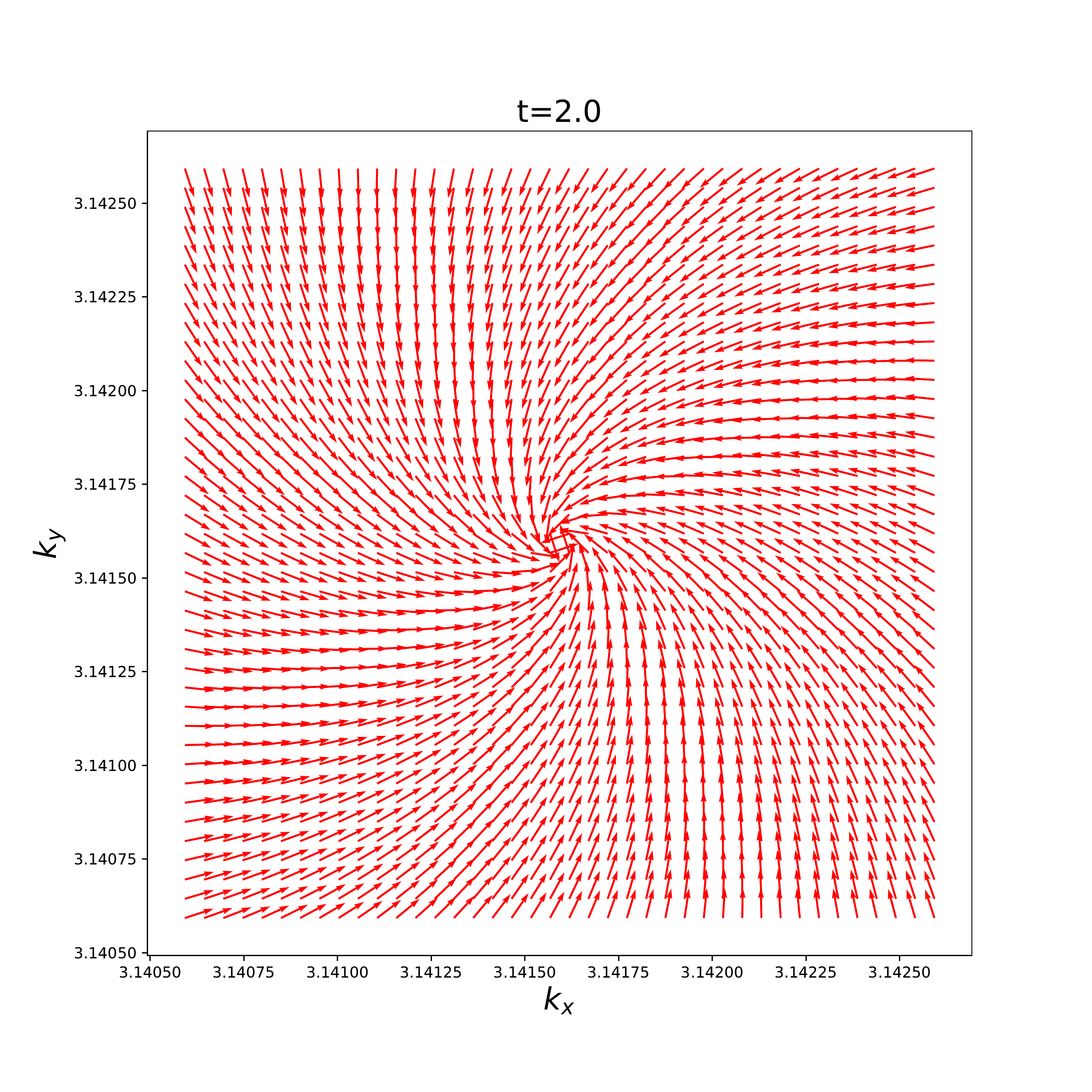}
\end{minipage}
\begin{minipage}{0.49\hsize}
            \centering
    \includegraphics[width=\hsize]{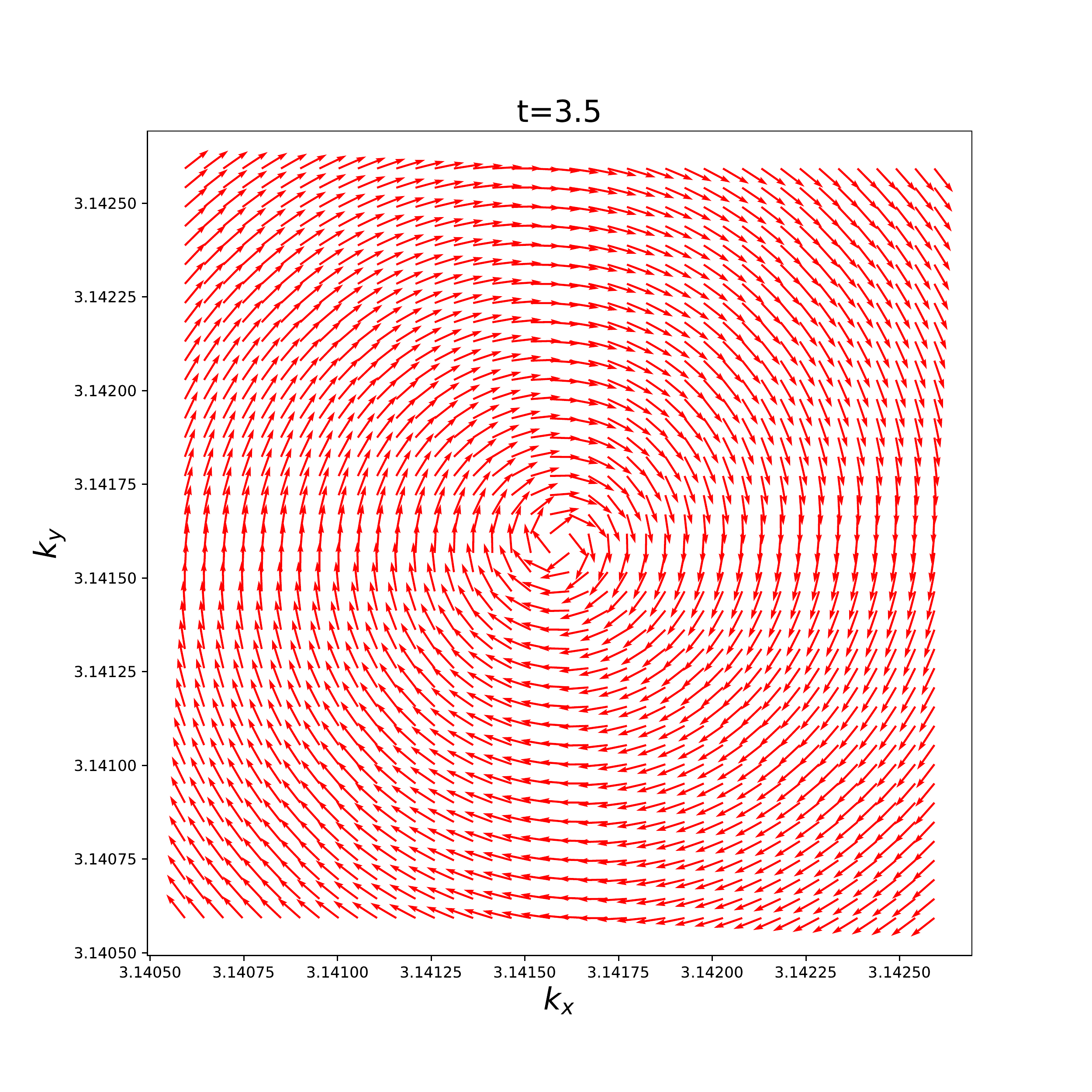}
\end{minipage}
\begin{minipage}{0.49\hsize}
            \centering
    \includegraphics[width=\hsize]{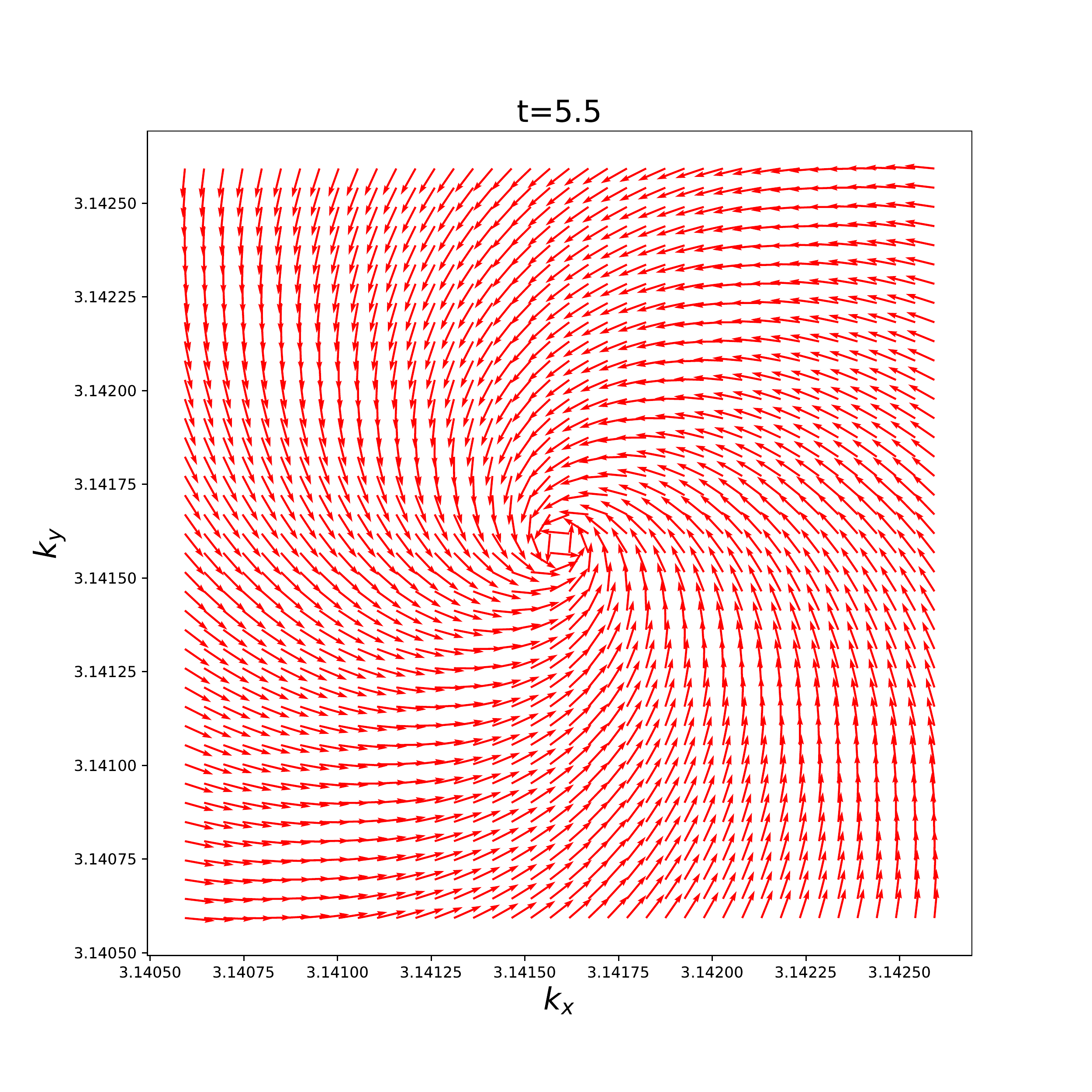}
\end{minipage}
\begin{minipage}{0.49\hsize}
            \centering
    \includegraphics[width=\hsize]{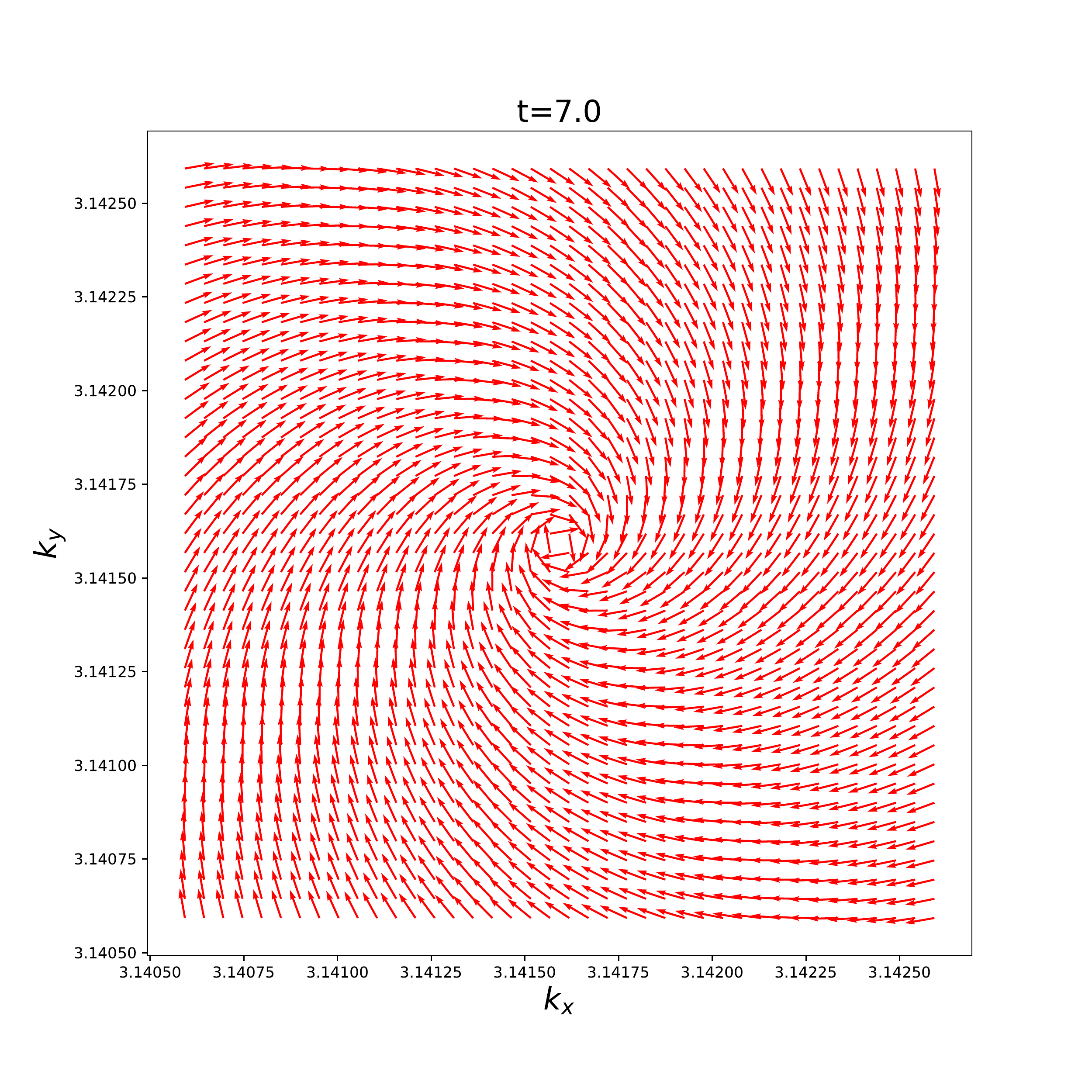}
\end{minipage}
\begin{minipage}{0.49\hsize}
            \centering
    \includegraphics[width=\hsize]{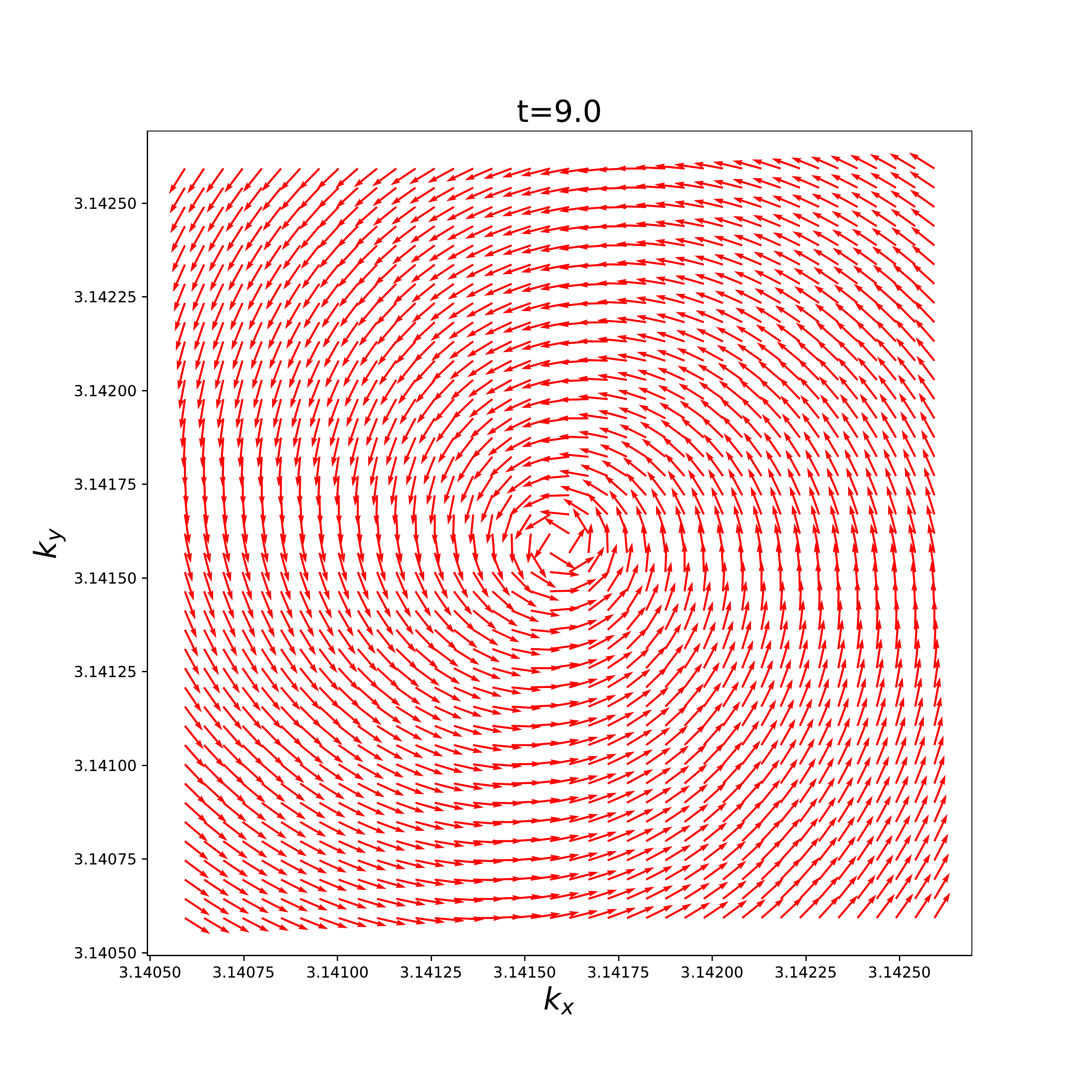}
\end{minipage}
    \caption{Time evolution of the vortex after quench at $k_x=k_y=\pi$.}
    \label{fig:local}
\end{figure*}

\begin{figure*}
\begin{minipage}{0.49\hsize}
    \centering
    \includegraphics[width=\hsize]{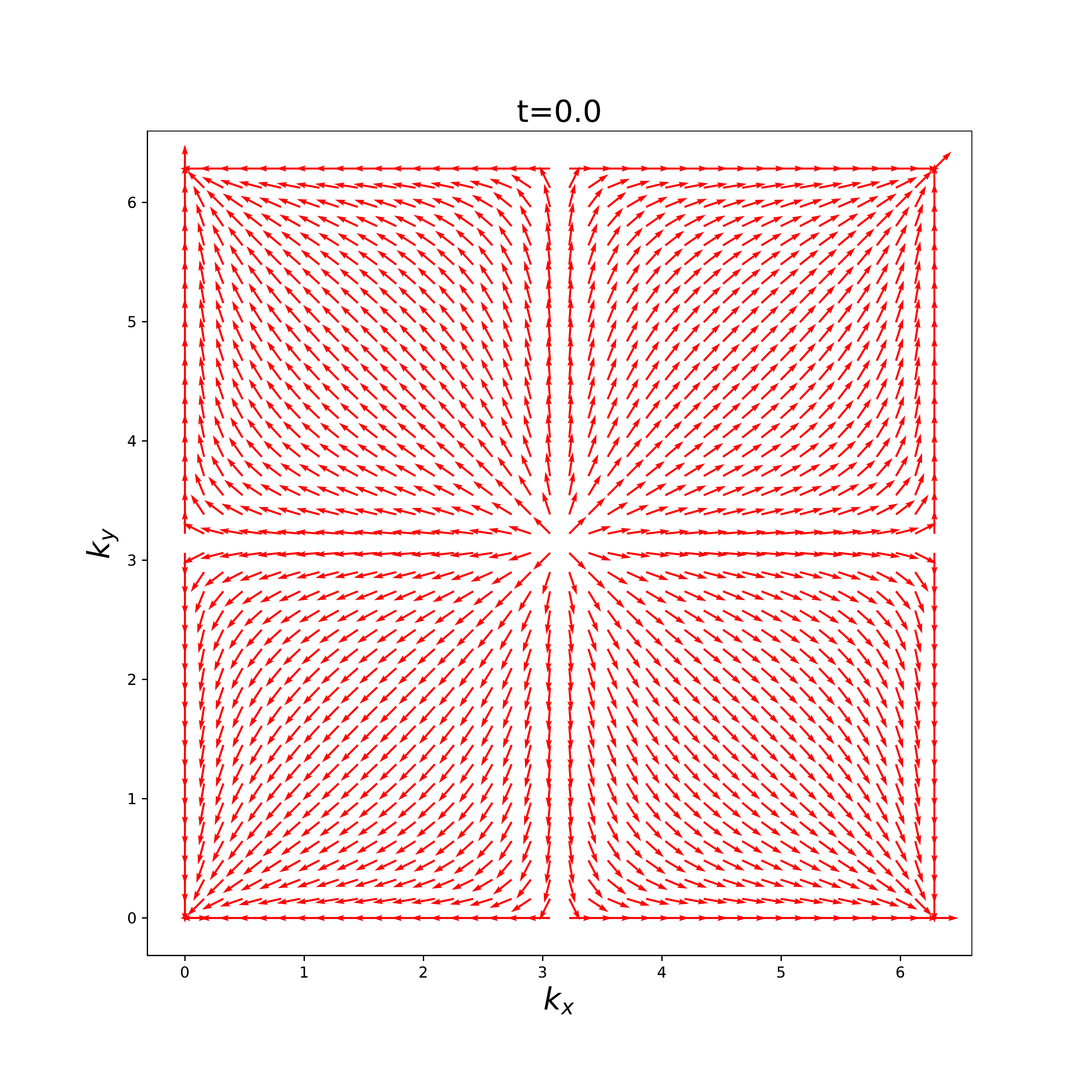}
\end{minipage}
\begin{minipage}{0.49\hsize}
    \centering
    \includegraphics[width=\hsize]{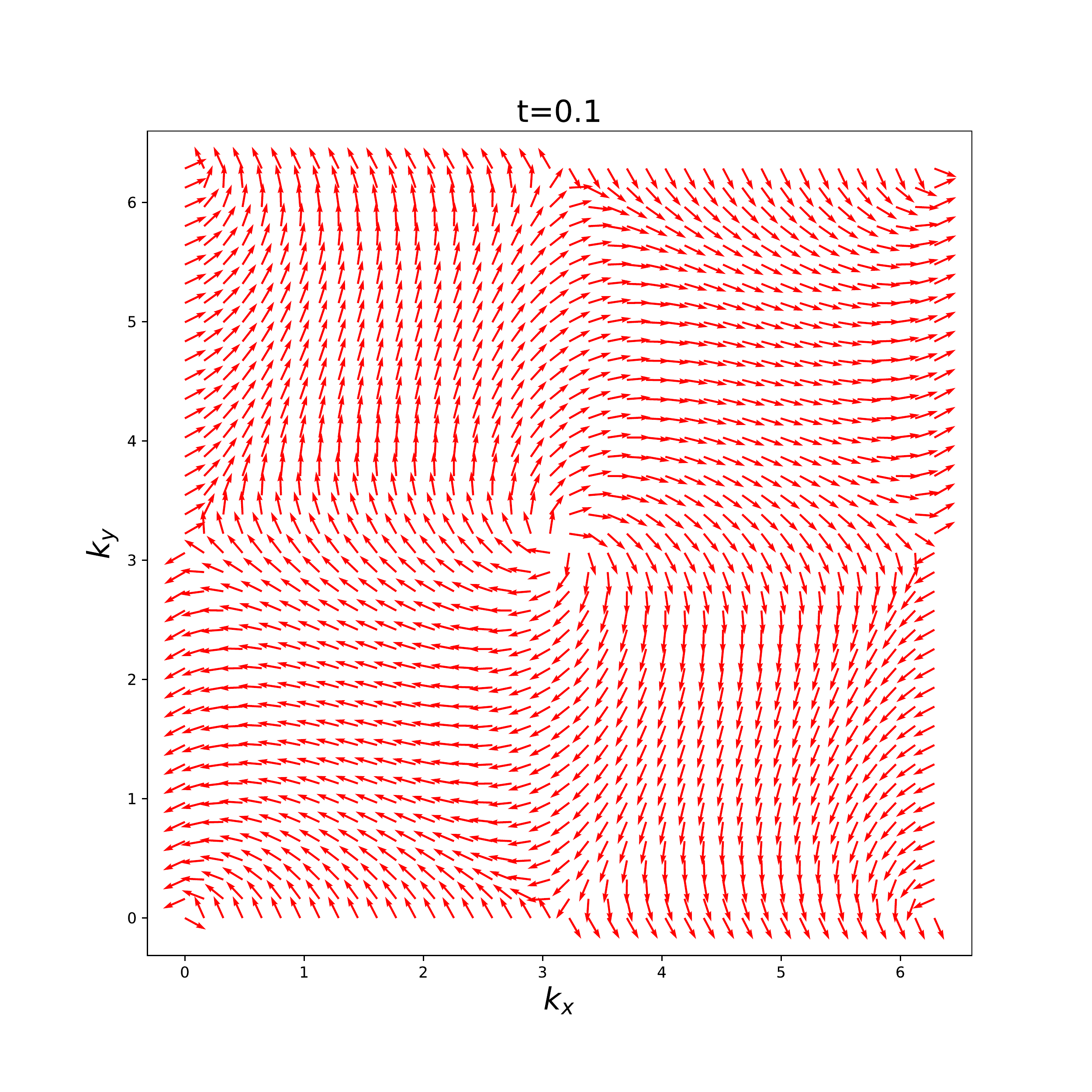}
\end{minipage}
\begin{minipage}{0.49\hsize}
    \centering
    \includegraphics[width=\hsize]{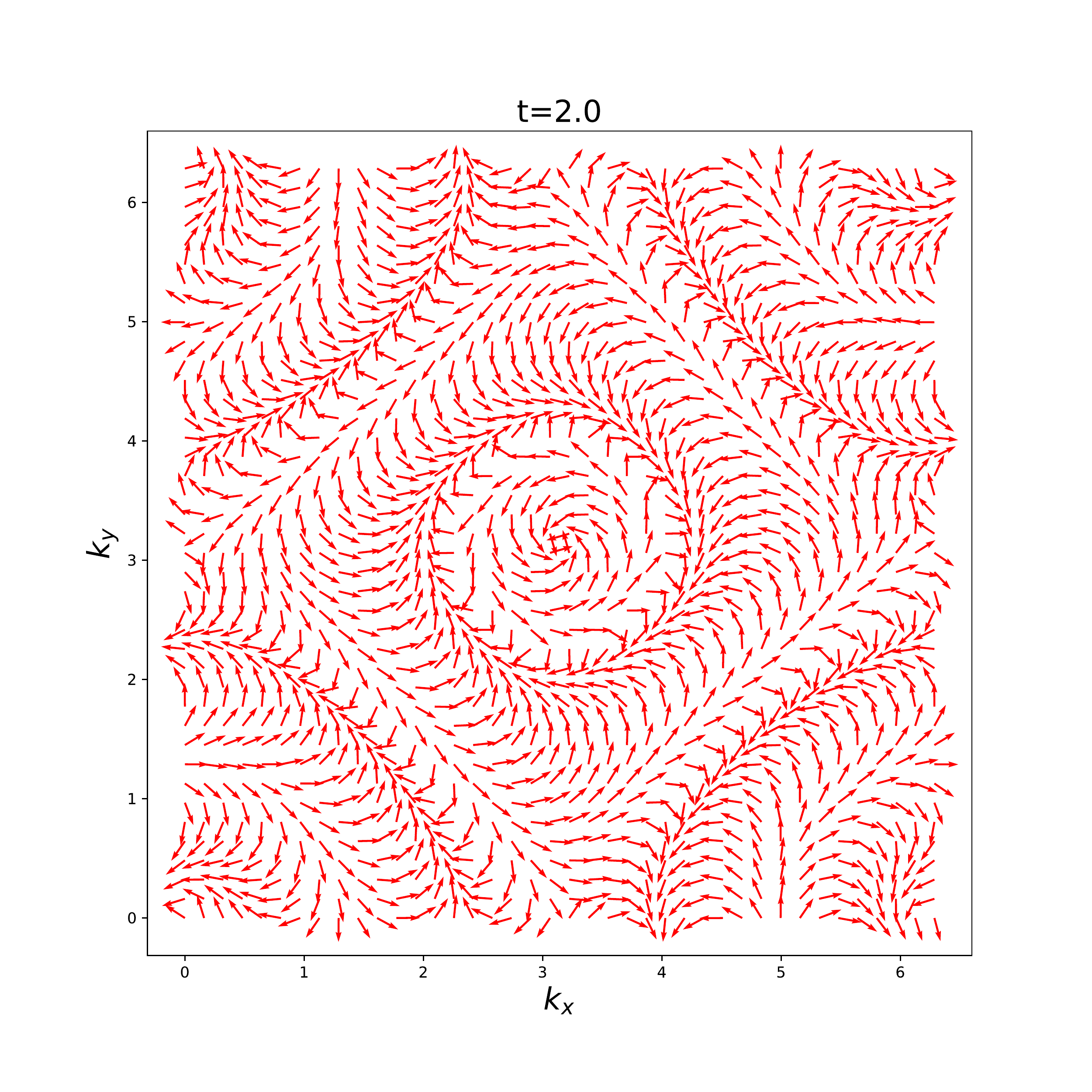}
\end{minipage}
\begin{minipage}{0.49\hsize}
    \centering
    \includegraphics[width=\hsize]{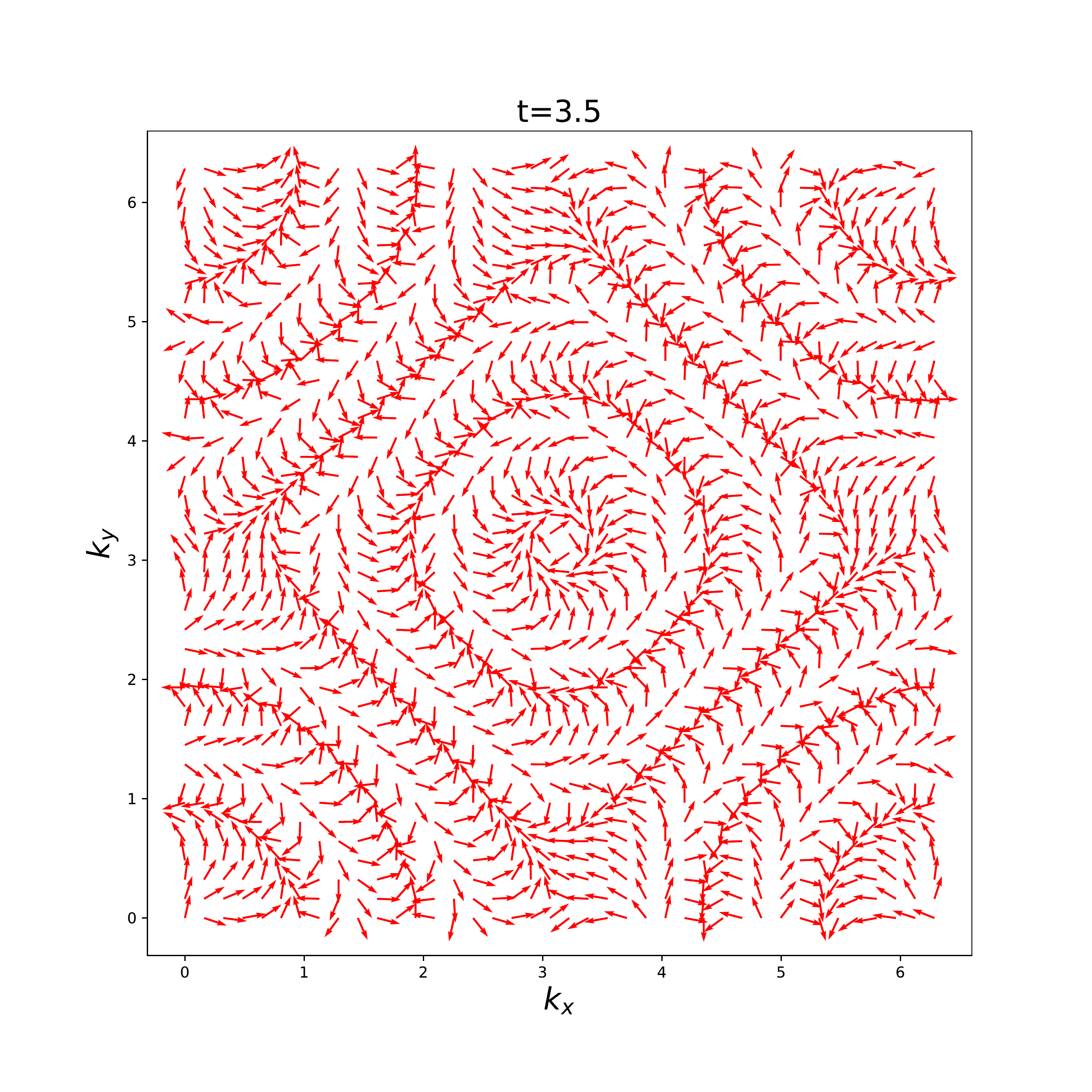}
\end{minipage}
\begin{minipage}{0.32\hsize}
    \centering
    \includegraphics[width=\hsize]{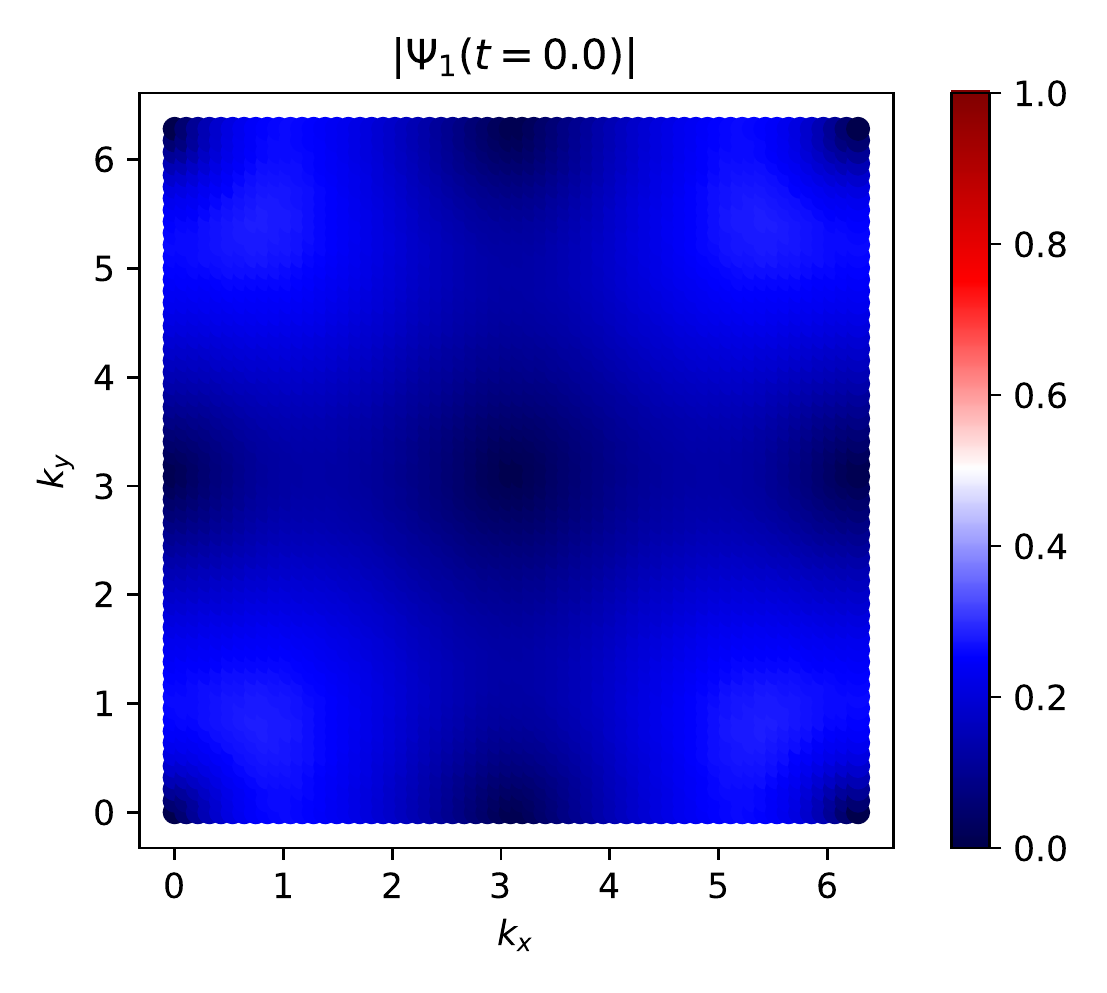}
\end{minipage}
\begin{minipage}{0.32\hsize}
    \centering
    \includegraphics[width=\hsize]{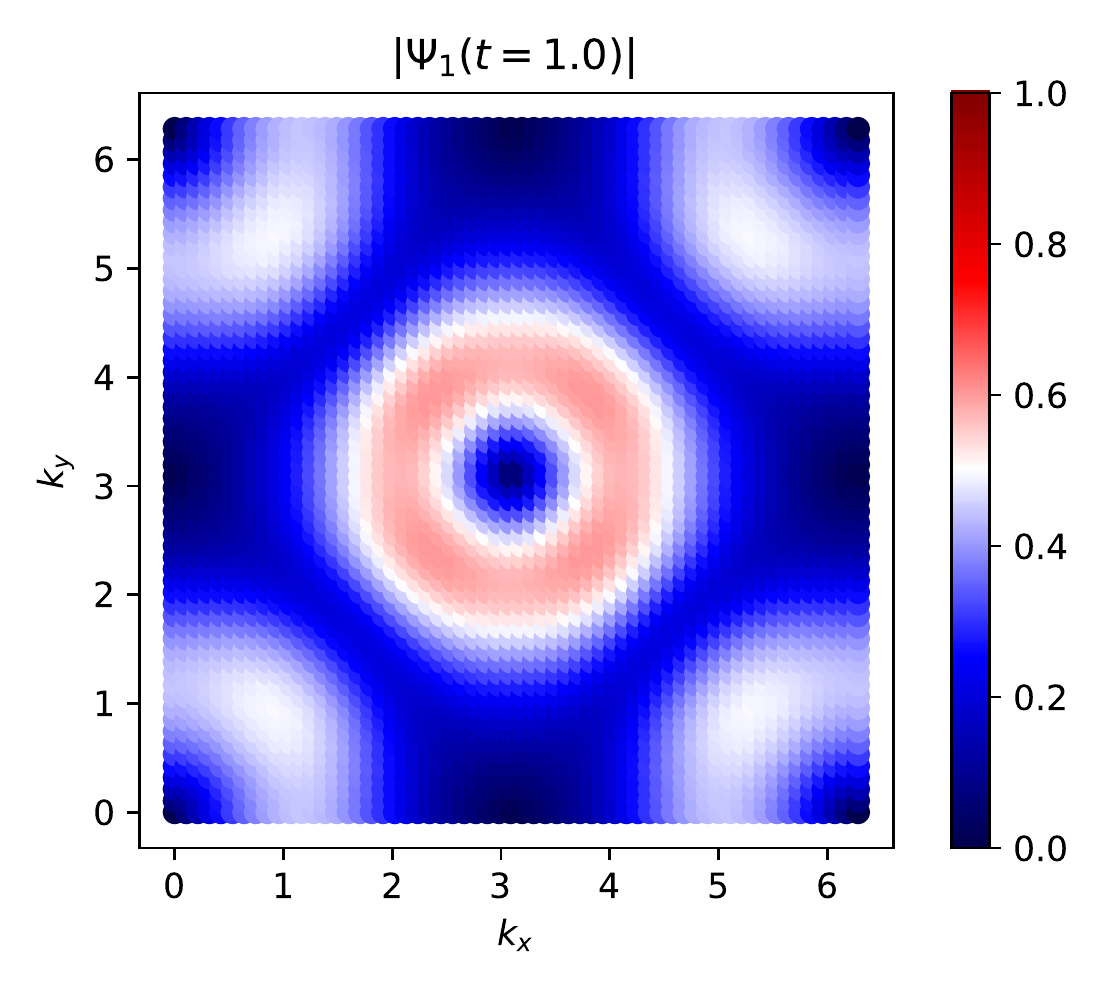}
\end{minipage}
\begin{minipage}{0.32\hsize}
    \centering
    \includegraphics[width=\hsize]{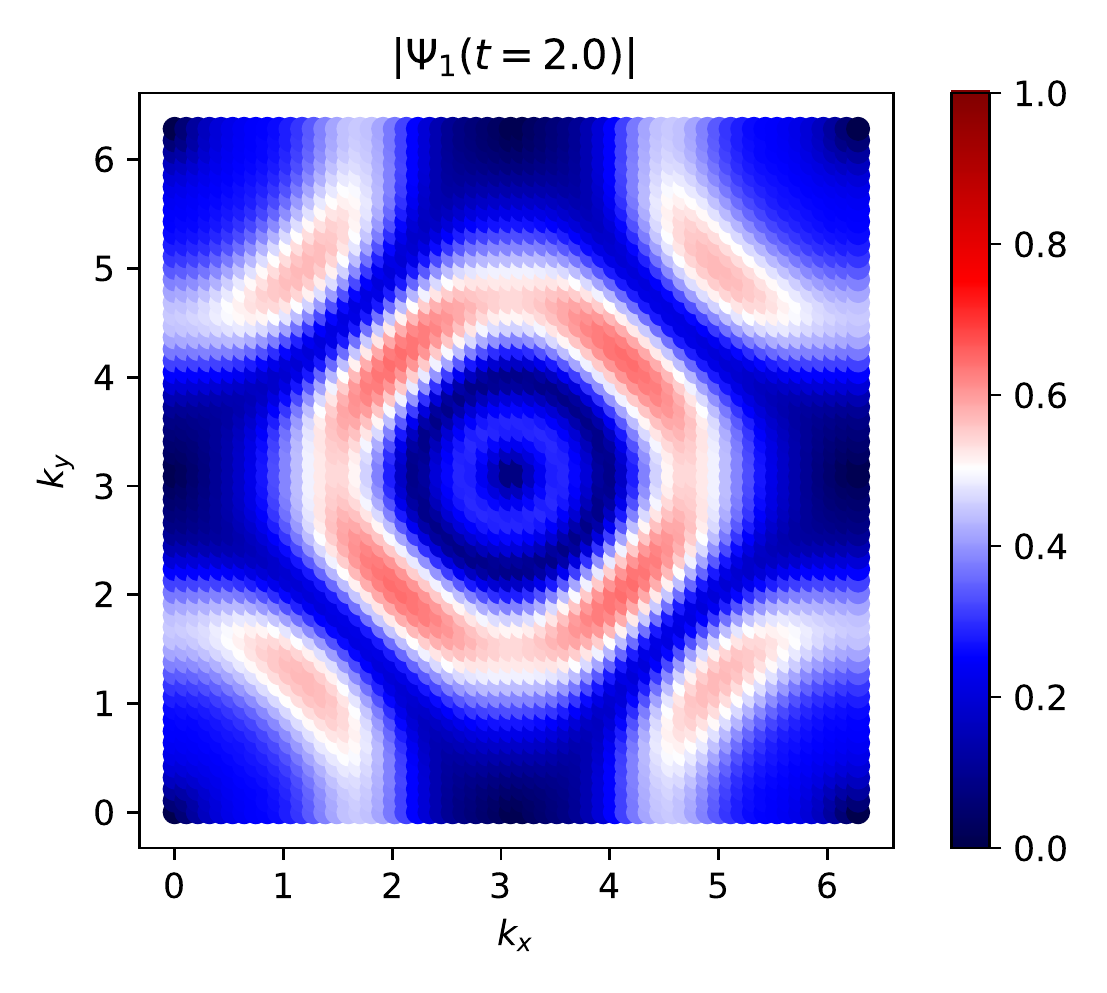}
\end{minipage}
    \caption{Time evolution of the vortex after quench. [Bottom] Time evolution of the density corresponding to the first component of the state after quench.}
    \label{fig:vortex}
\end{figure*}
\subsection{Vortices and topology}

In general the Berry connection possesses several vortices in the magnetic Brillouin zone. Here we are going to discuss the  definition of the Chern number in this more general case. 

The two Bloch states $\ket{\Psi^A}, \ket{\Psi^B}$ defined in different gauges are related by
\begin{equation}
    \ket{\Psi^B(t,k)}=e^{i\xi_{BA}(t,k)}\ket{\Psi^A(t,k)}. 
\end{equation}
For an arbitrary Bloch state $\ket{\Psi}$, 
we choose a complex number $\xi_A$ such that the second component of $\Psi^A$ is strictly positive:
\begin{align}
\begin{aligned}
\ket{\Psi^A}&=e^{-i\text{Arg}\,\xi_A}\ket{\Psi},\quad  
(\Psi^A)_2>0.
\end{aligned}
\end{align}
This is possible everywhere except for zeros of $\xi_A$. We will assume that these zeros are isolated, and define  $S_i,(i=1,\cdots,N)$ as small disks around each of them. Around those zeros, we define $\Psi^B$ by 
\begin{align}
\begin{aligned}
\ket{\Psi^B}&=e^{-i\text{Arg}\,\xi_B}\ket{\Psi},\quad   
(\Psi^B)_1>0,
\end{aligned}
\end{align}
again choosing the phase of $\xi_B$ such that the first component of $\Psi^B$ is positive.
Then the gauge transformation is given by 
\begin{equation}
    \xi_{BA}=\text{Im log}\frac{\xi_A}{\xi_B}. 
\end{equation}
Therefore, due to the Stokes theorem, given $S=\bigcup_{i=1}^NS_i$, the Chern number associated with the Berry connection can be written as 
\begin{align}
\begin{aligned}
    C(t)&=\frac{1}{2\pi}\int_Sd^2k d\mathcal{A}^B+\frac{1}{2\pi}\int_{T_{\text{BZ}}\setminus S}d^2k d\mathcal{A}^B,\\
    &=\frac{1}{2\pi}\sum_{i=1}^N\oint _{\partial S_i}dk\cdot(\mathcal{A}^B-\mathcal{A}^A)\\
    &=\sum_{i=1}^N\frac{1}{2\pi}\oint_{\partial S_i}dk \cdot\nabla\xi_{BA}(t,k). 
\end{aligned}    
\end{align}
Consequently the total Chern number is related to the total vorticity of Bloch state.

\subsection{Topological quench protocol}\label{sec:memory}

In our quench protocol, we put the system in the ground state at an initial value of $m$, and then at time $t=0$ abruptly change $m$ to a new value of $m'$. We then measure the Loschmidt amplitude, which is a measure of fidelity between two quantum states in a quench process:
\begin{align}
    L_{m\to m'}(t)&=\bra{\psi_{n,m}}e^{-it H_{m'}}\ket{\psi_{n,m}},
\end{align}
where $\psi_{n,m}$ is the $n$-th eigenstate of the Hamiltonian $H_m$ with mass $m$. The Loschmidt amplitude has been used for detecting dynamical quantum phase transitions in ~\cite{2020PhRvB.102j4305H,2018arXiv180807885Z}. 

For $m>2$ (when there are no vortices in the ground state), we use $\ket{\Psi_\text{in}}=\ket{\Psi^B}$ for the initial state and consider time evolution of vortices after quench, when $|m'|<2$. For $|m'|<2$ we use the ground state $\ket{\Psi^B}$ or $\ket{\Psi^A}$ around a circle at $(\pi,\pi)$. The subsequent time evolution for $|m'|<2$ described by  $e^{-it H_{m'}}\ket{\Psi_\text{in}}$ corresponds to the gauge transformation.

The local vortex evolution is shown in Fig.~\ref{fig:local}, and the time evolution of the vector field over the entire Brillouin zone is shown in Fig.~\ref{fig:vortex}. As it is clear from Fig.~\ref{fig:local}, the direction of the vortex at $k_x=k_y=\pi$ changes periodically. We can assign the $\mathbb{Z}_2$-index to the vortex based on its rotation direction : clockwise (-1) or anticlockwise (+1).  Fig.~\ref{fig:vortex} exhibits the vortex at $(k_x,k_y)=(\pi,\pi)$ and Fig.~\ref{fig:vortex} [Bottom] shows the time evolution of the ground state density (lower band). Fig.~\ref{fig:pi} and \ref{fig:zero} show the variation of Loschmidt amplitude and vorticity at $k_x=k_y=\pi$ and 0, respectively.

\begin{figure}[H]
    \centering
    \includegraphics[width=\hsize]{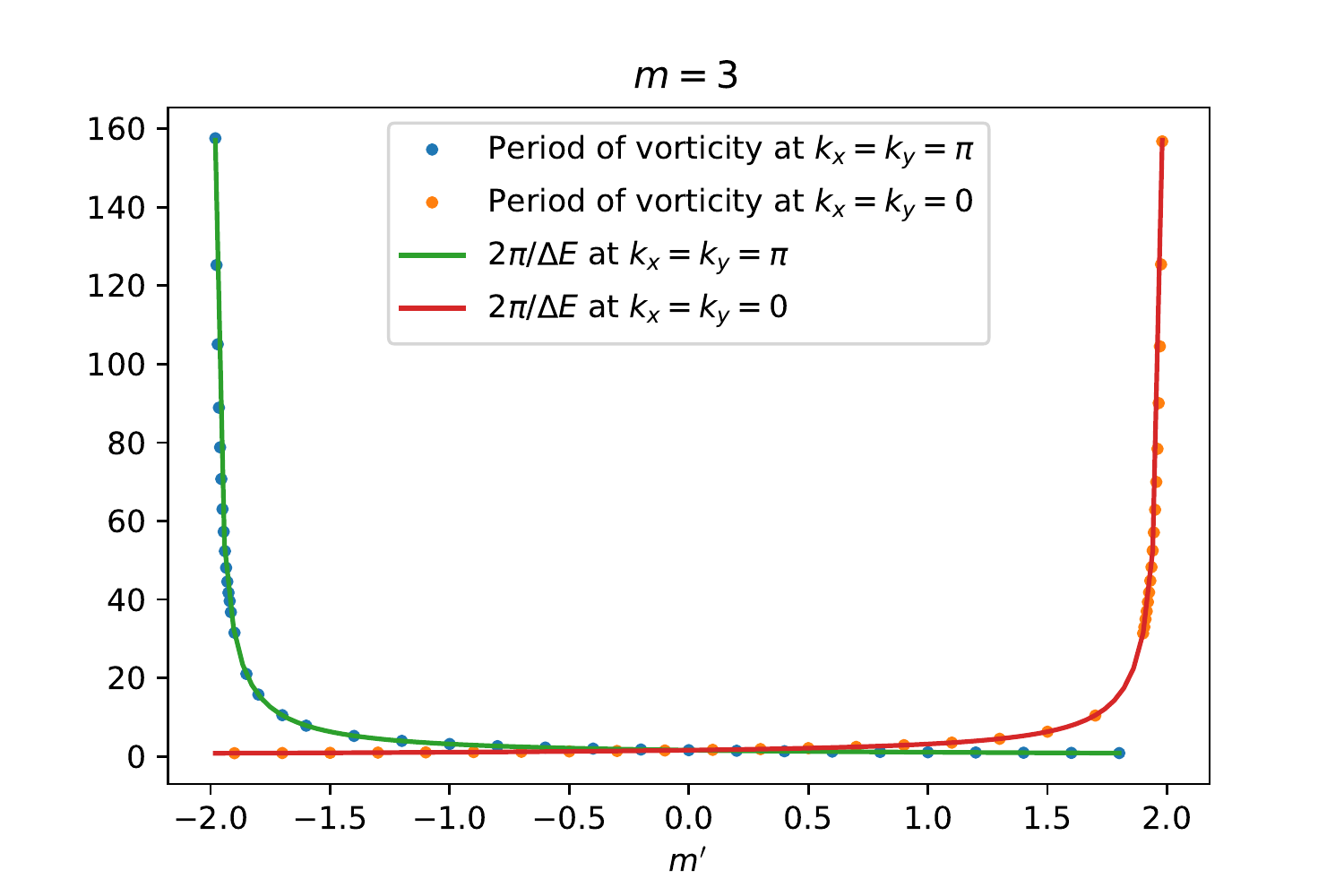}
    \caption{$m'$-dependence of the period of vorticity oscillations at $k_x=k_y=\pi$. The initial mass is set to $m=3$.}
    \label{fig:VorticityPeriod}
\end{figure}

\begin{figure*}
\begin{minipage}{0.32\hsize}
    \centering
    \includegraphics[width=\hsize]{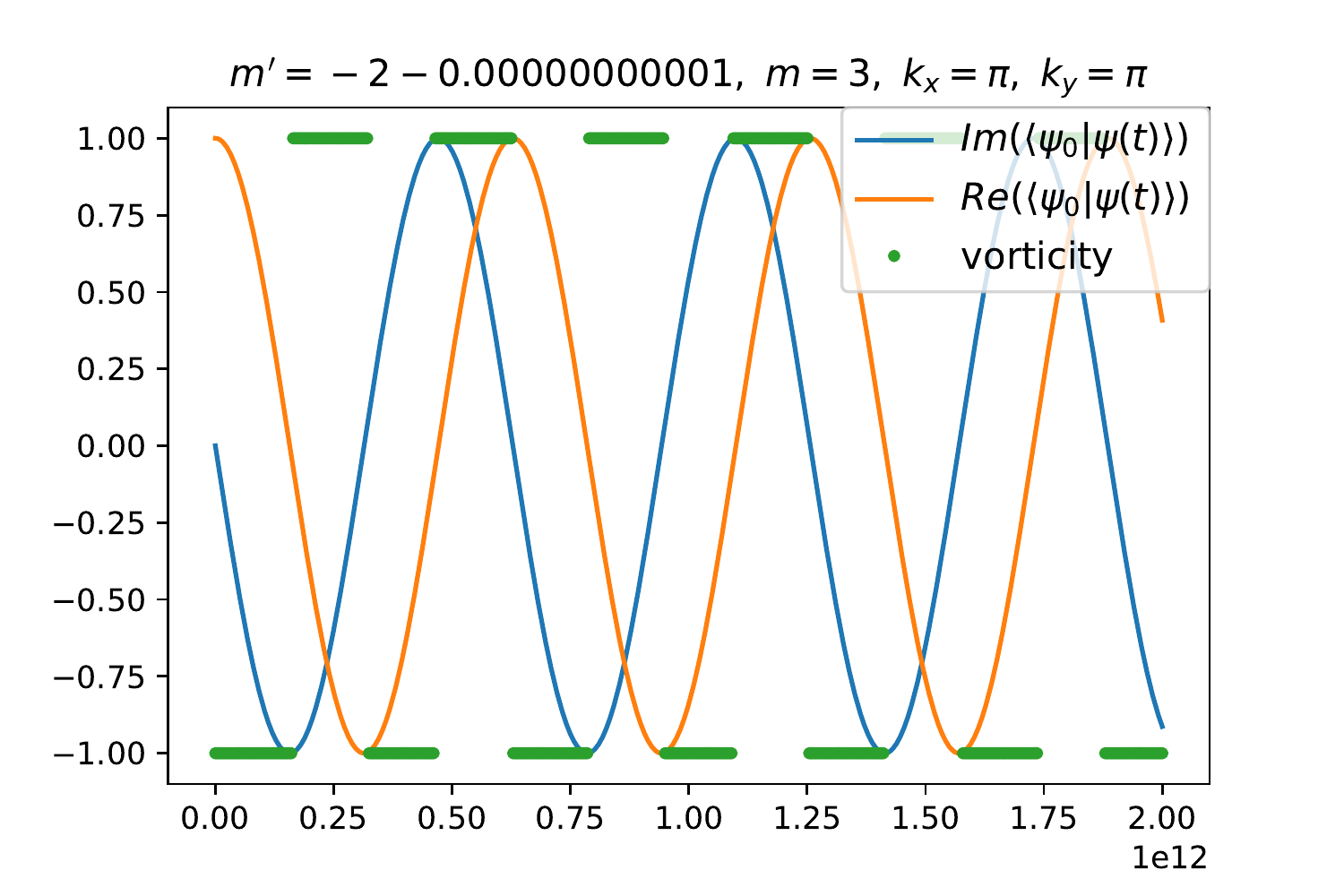}
\end{minipage}
\begin{minipage}{0.32\hsize}
    \centering
    \includegraphics[width=\hsize]{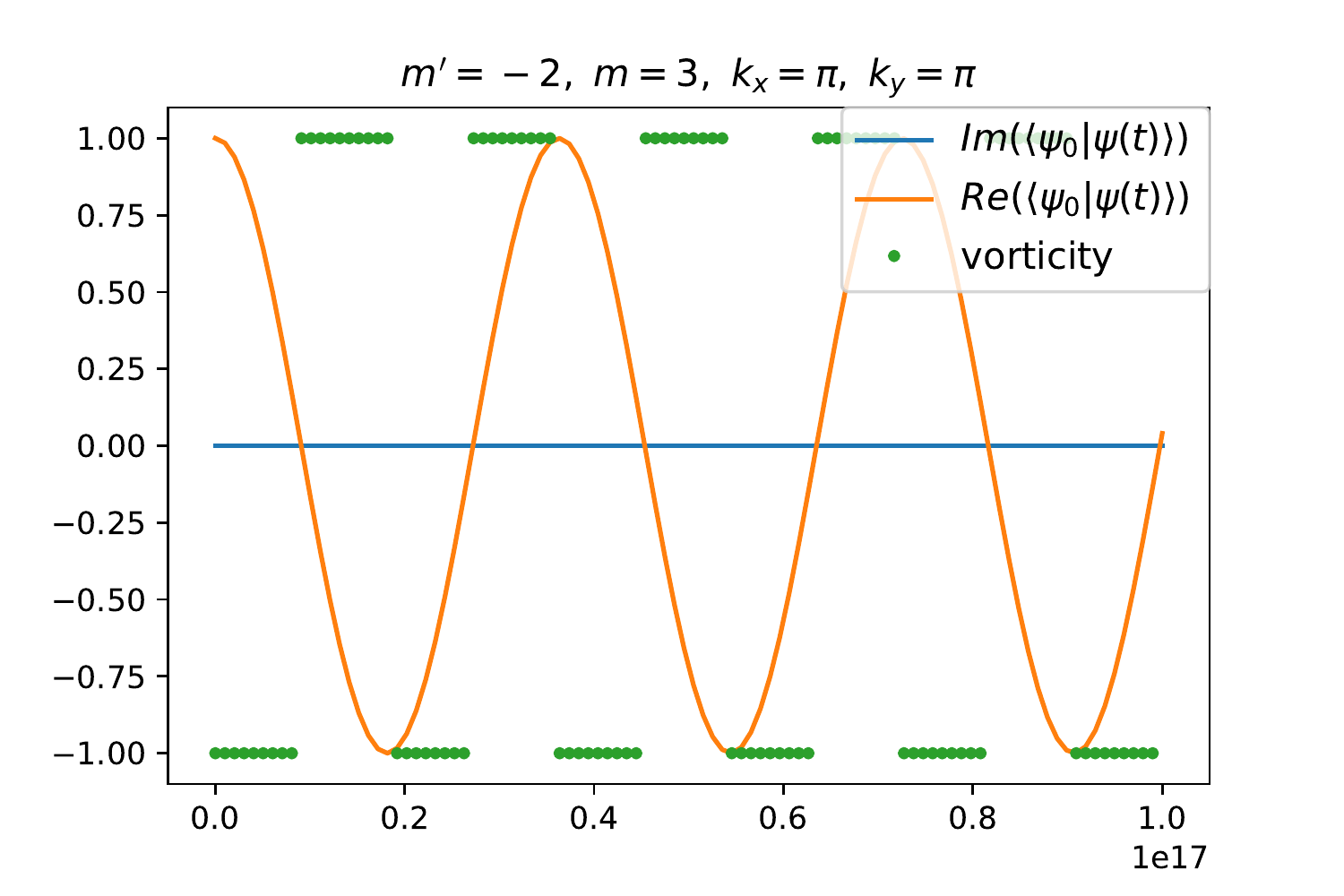}
\end{minipage}
\begin{minipage}{0.32\hsize}
            \centering
    \includegraphics[width=\hsize]{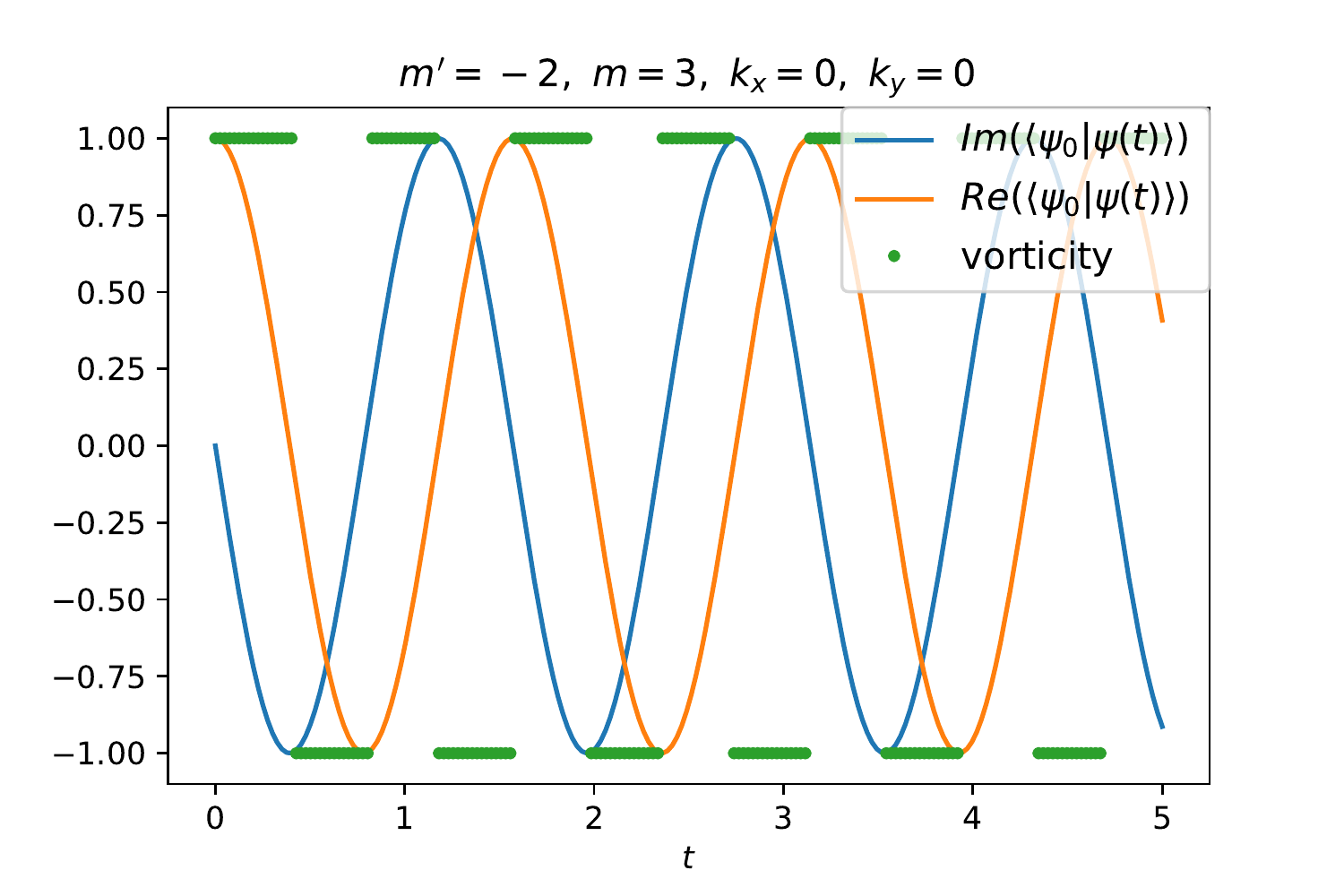}
\end{minipage}
    \caption{Vorticity at $m=-2,k_x=k_y=\pi$ [Left and Middle] and at $m=-2,k_x=k_y=0$.}
    \label{fig:vorticity_m-2}
\end{figure*}

\begin{figure*}
\begin{minipage}{0.49\hsize}
        \centering
    \includegraphics[width=\hsize]{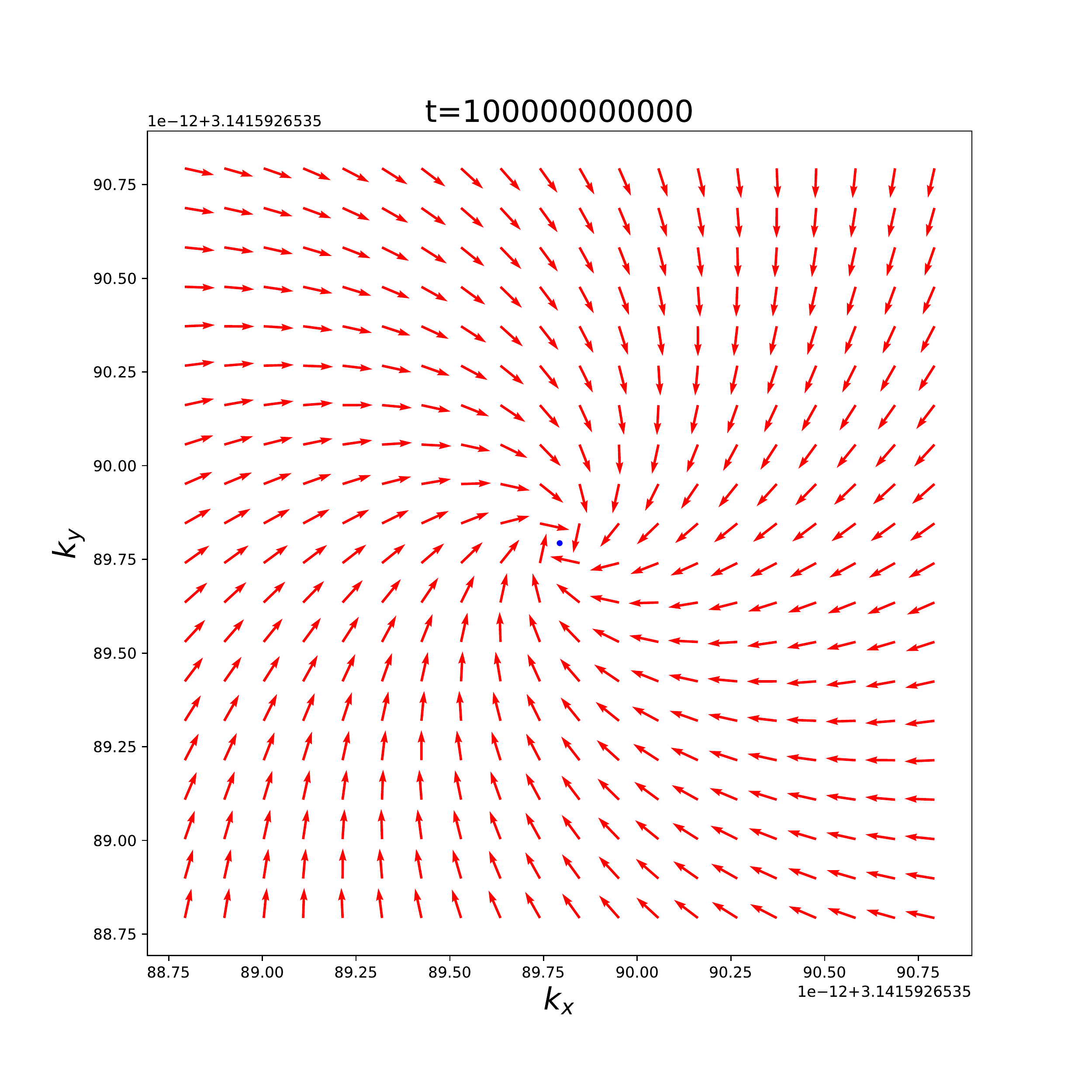}
\end{minipage}
\begin{minipage}{0.49\hsize}
        \centering
    \includegraphics[width=\hsize]{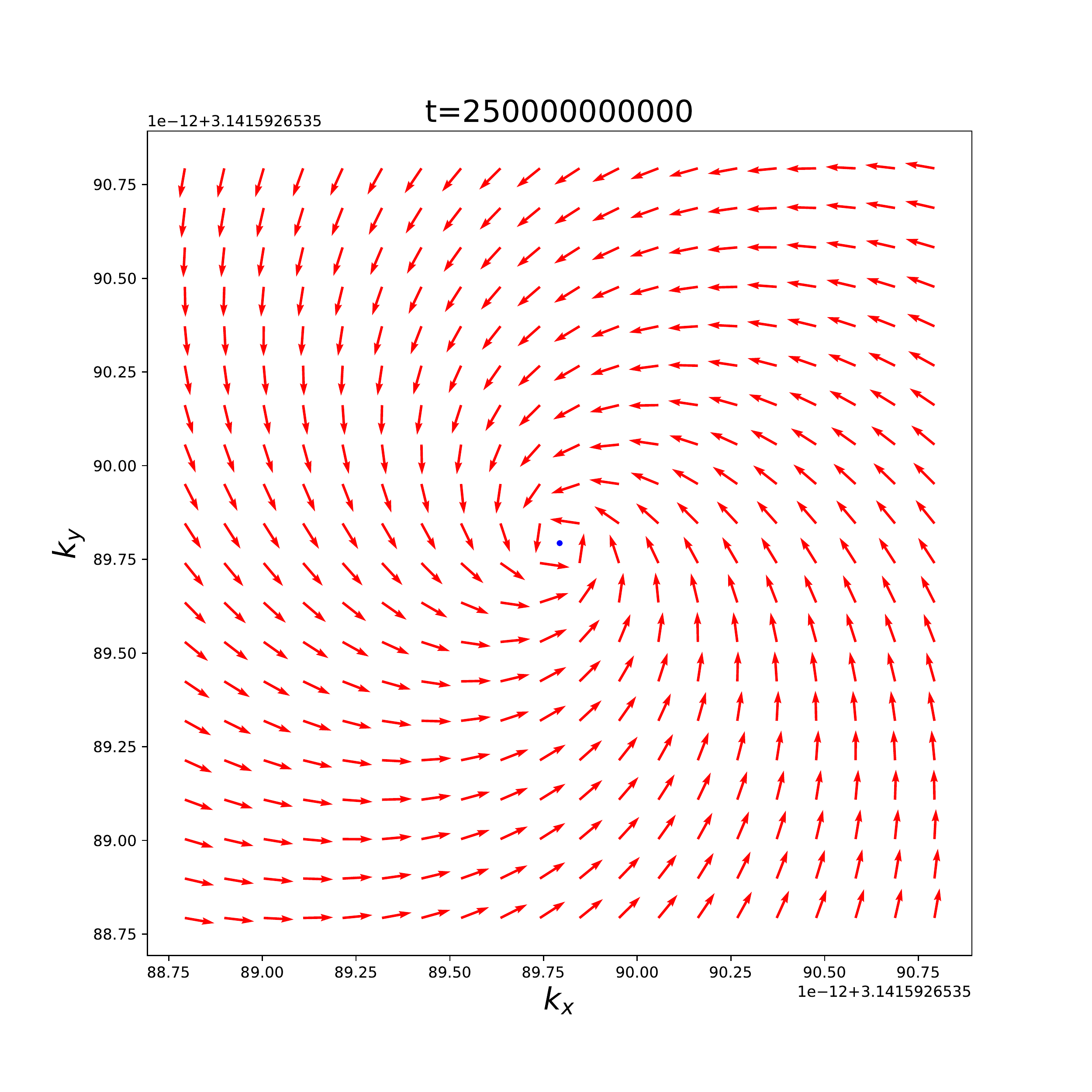}
\end{minipage}
\begin{minipage}{0.49\hsize}
        \centering
    \includegraphics[width=\hsize]{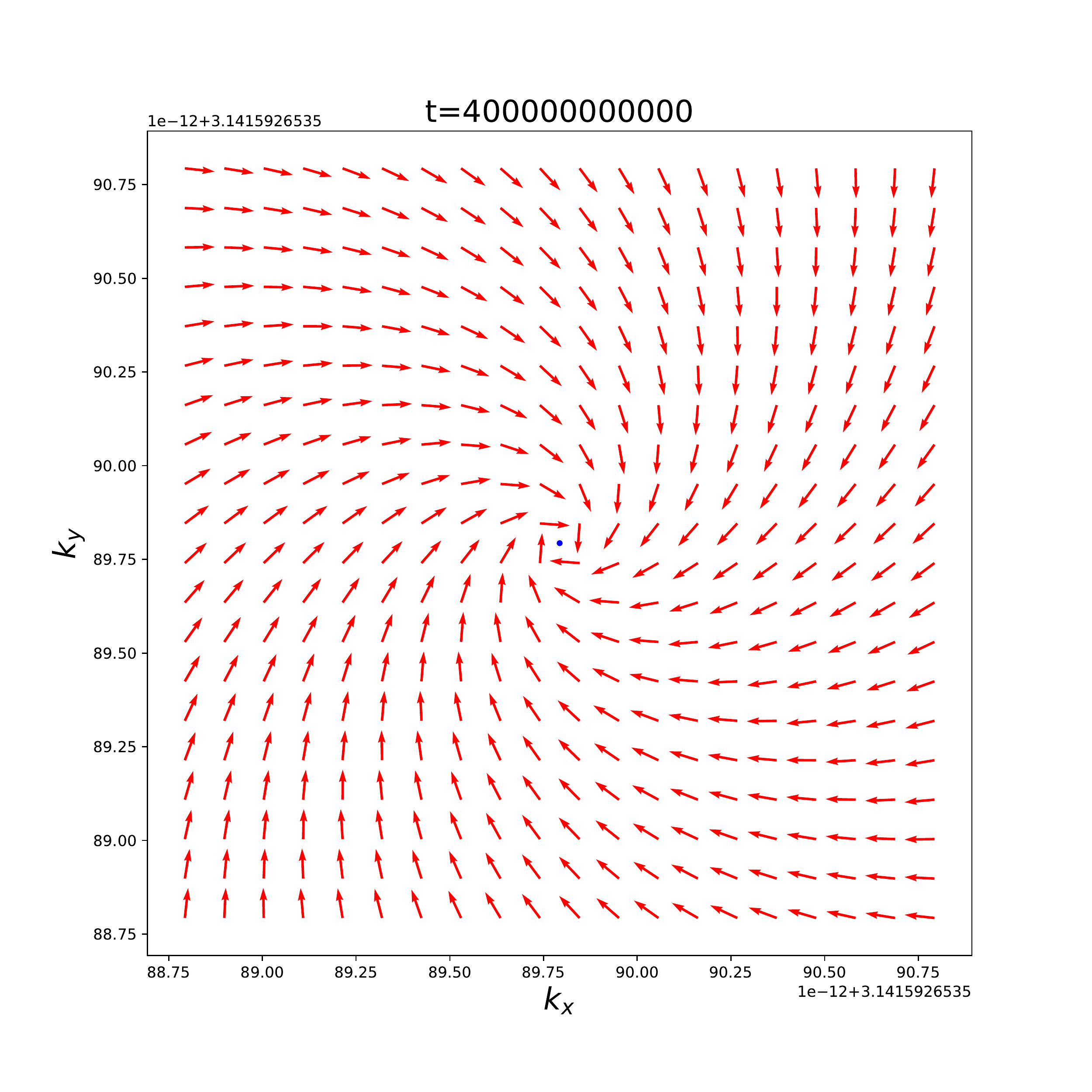}
\end{minipage}
\begin{minipage}{0.49\hsize}
        \centering
    \includegraphics[width=\hsize]{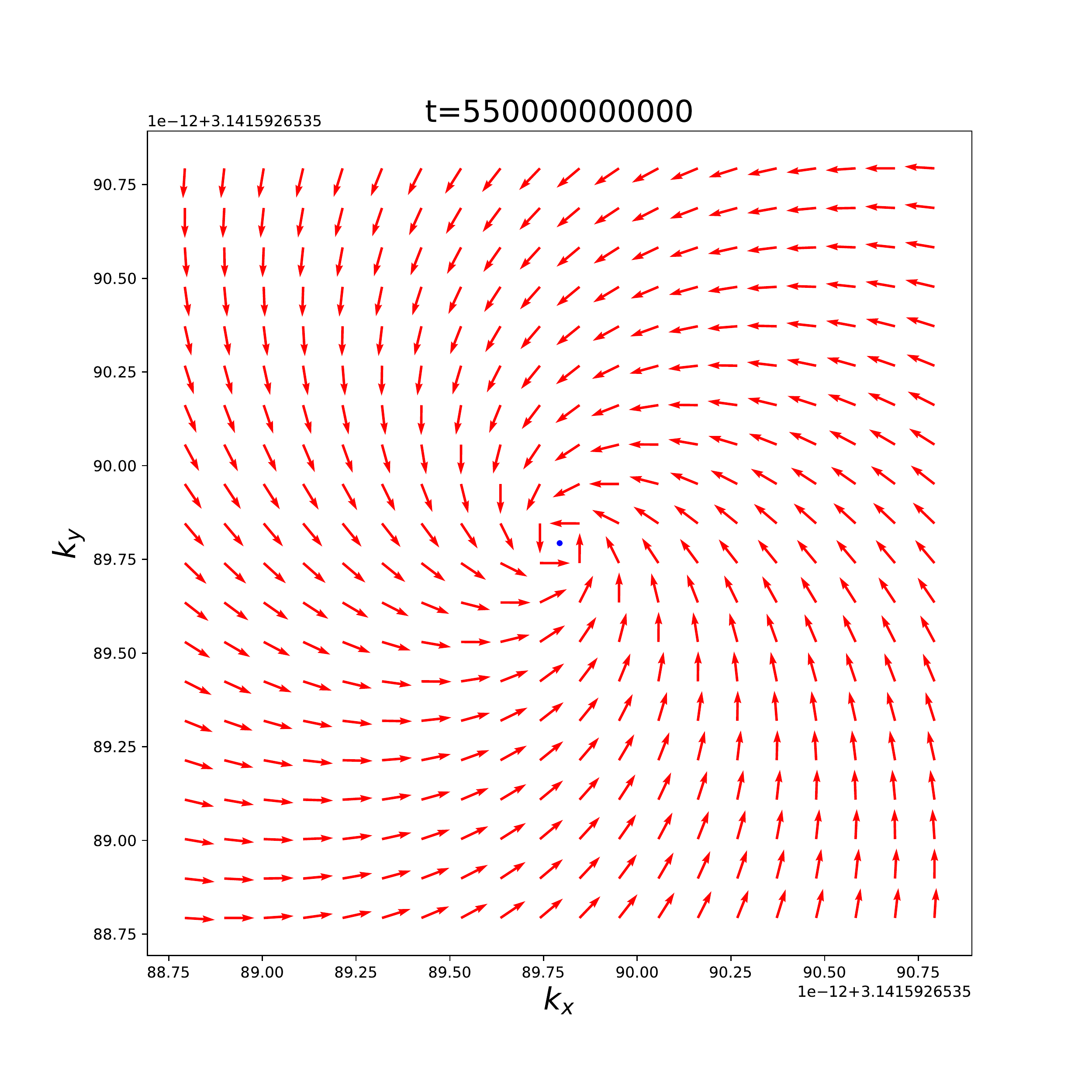}
\end{minipage}
    \caption{Vortex at $m=-2-0.00000000001$ at $k_x=k_y=\pi$}
    \label{fig:vortex_m-2}
\end{figure*}

Our numerical calculations reveal that the time at which the sign of the real or imaginary part of the Loschmidt amplitude changes coincides with the time at which the vorticity changes, and that plateaus with constant vorticity appear periodically in time. Oscillations between vortices and antivortices have been reported before \cite{2007PhRvA..76b1605W}, with the period $2\pi/\Delta E$, where $\Delta E$ is the energy difference between the two bands. Fig.~\ref{fig:VorticityPeriod} clearly shows that this law is perfectly valid in our case (see also Fig.~\ref{fig:energy}). The vortices at $k_x=k_y=\pi$ and $0$ have different periods, indicating that quantum memories with different periods are created in the Brillouin zone. The change in vorticity in the vicinity of $m=-2$ is shown in Fig.~\ref{fig:vorticity_m-2}, where its period is very long for $k_x=k_y=\pi$ (see also Fig.~\ref{fig:vortex_m-2}), while the period for $k_x=k_y=0$ is short. These results indicate the scaling of the period with the inverse band gap.

\section{Discussion and outlook}\label{sec:discussion}

Quantum memory records a quantum state for later retrieval. Various quantum memories have been theoretically proposed and are being actively explored experimentally~\cite{lvovsky2009optical,hedges2010efficient,wang2019efficient,julsgaard2004experimental,specht2011single,dennis2002topological}. Our proposal provides a simple topological quantum memory that can be realized with a single qubit operator. The system of oscillating vortices and antivortices records information about the amplitude of an electromagnetic wave (encoded through the magnetic flux parameter $m$ of the Hamiltonian) through the period of oscillations, and about its phase. Therefore, both the phase and the amplitude of an electromagnetic wave are recorded in the Quantum Hall state, which thus makes it a quantum memory. Moreover, the period of oscillations varies in a very wide interval depending on the external magnetic flux. 

Readout of the quantum state recorded in this ``vortical quantum memory" can be realized optically since the reflection and transmission of electromagnetic waves will be affected by the presence of oscillating vortices and the resulting from them oscillating Quantum Hall conductivity. In fact, the two-band model used in this study has already been implemented in many methods, such as photonic crystals~\cite{2016arXiv161102373N}
microwave resonators~\cite{peterson2018quantized}, superconducting qubit~\cite{NIU20211168}, phononic metamaterials~\cite{serra2018observation}, and its topological properties have been investigated. Clearly, the theory of this electromagnetic readout should be developed quantitatively. We leave this for a future investigation. 

It will also be interesting to investigate the possibility of realizing quantum memory in other traditional class~\cite{PhysRevLett.61.2015,PhysRevLett.95.146802,PhysRevB.85.075125} and new class of topological models, e.g. those described in~\cite{2020PhRvL.125e3901Y,2019PhRvR...1c3079J,2020arXiv200805489M,2021arXiv210413314I,2021arXiv210710586A}.

\if{
\begin{figure*}
\begin{minipage}{0.49\hsize}
            \centering
    \includegraphics[width=\hsize]{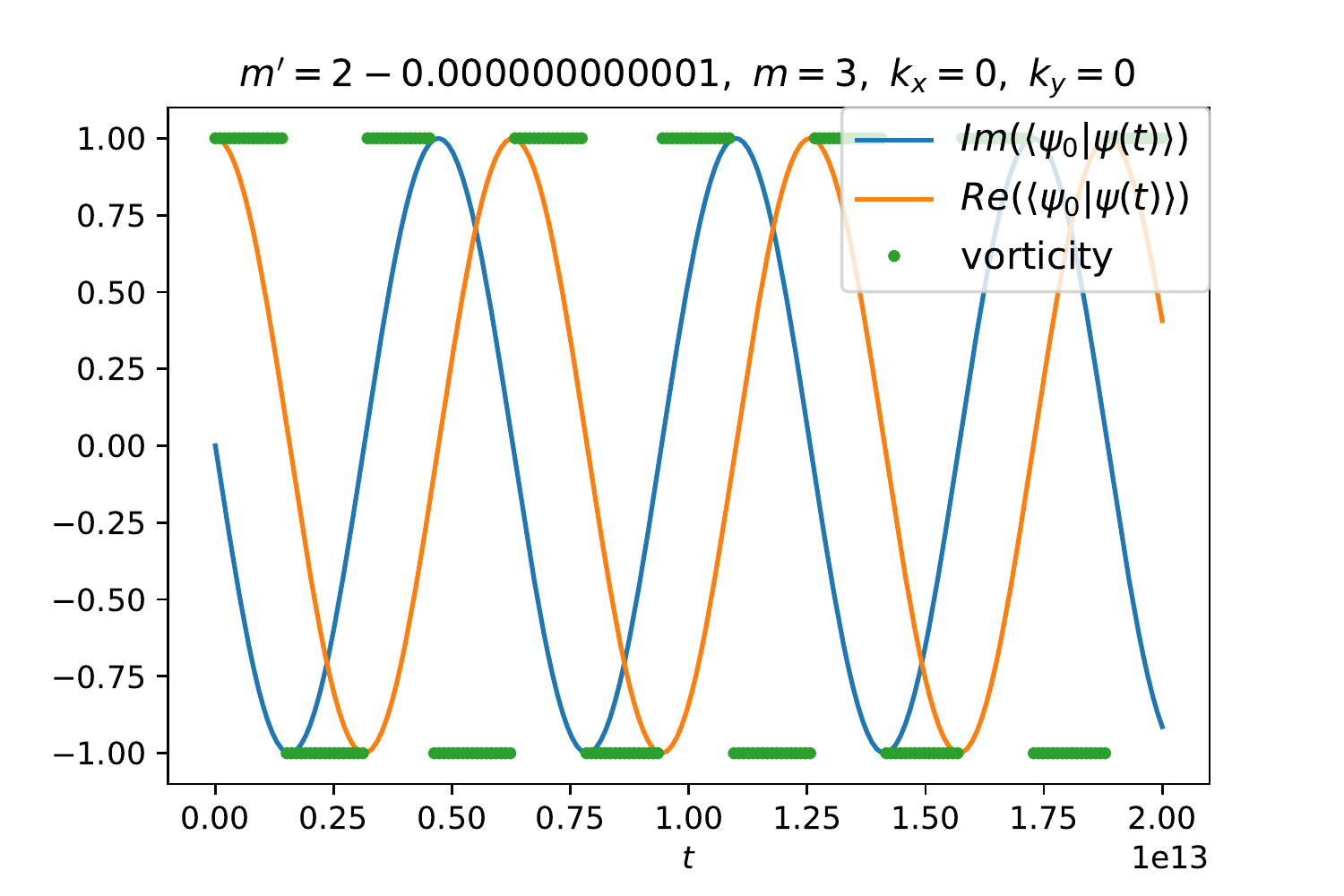}
\end{minipage}
    \caption{Vorticity at $m=2,k_x=k_y=0$ }
    \label{fig:kx0ky0}
\end{figure*}
}\fi

\section*{Acknowledgement}
This work was supported by PIMS Postdoctoral Fellowship Award (K. I.) and by the U.S. Department of Energy under awards DE-SC0012704 (D. K. and Y. K.) and DE-FG88ER40388 (D. K.). The work on numerical simulations was supported by the U.S. Department of Energy, Office of Science National Quantum Information Science Research Centers under the ``Co-design Center for Quantum Advantage" award. K.I. thanks Yoshiyuki Matsuki for useful communications. 

\bibliographystyle{utphys}
\bibliography{ref}
\end{document}